\documentclass[draft,onecolumn,12pt]{IEEEtran}

\usepackage[noadjust]{cite}
\usepackage{graphicx}
\usepackage{subfigure}
\usepackage{textcomp,amssymb,amsmath}
\usepackage{makeidx,epsfig}
\usepackage{graphicx,subfigure}
\usepackage{algorithm}
\usepackage{color}

\newtheorem{lemma}{\sc Lemma}
\newtheorem{proposition}{\sc Proposition}

\newtheorem{remark}{\sc Remark}

\def\sinr{{\rm SINR}}

\def\tr{{\rm tr}}
\def\E{{\mathbf E}}
\def\calA{{\mathcal A}}
\def\calB{{\mathcal B}}
\def\calC{{\mathcal C}}
\def\calG{{\mathcal G}}

\def\calK{{\mathcal K}}

\def\calN{{\mathcal N}}
\def\calS{{\mathcal S}}

\def\bh{{\mathbf h}}

\def\bs{{\mathbf s}}

\def\bQ{{\mathbf Q}}

\psfull

\begin{document}

\title{Coordinated Multicasting with Opportunistic\\[-.3cm] User Selection in Multicell Wireless Systems}

\author{Y.-W. Peter Hong, Wei-Chiang Li, Tsung-Hui Chang, and Chia-Han Lee
\thanks{
Y.-W. P. Hong and W.-C. Li (emails: {\tt ywhong@ee.nthu.edu.tw} and {\tt
weichiangli@gmail.com}) are with the Institute of Communications Engineering,
National Tsing Hua University, Hsinchu, Taiwan. T.-H. Chang (email: {\tt tsunghui.chang@ieee.org}) is with the Department of Electronic and Computer Engineering, National Taiwan University of Science and Technology, Taipei, Taiwan.
C.-H. Lee (email: {\tt chiahan@citi.sinica.edu.tw}) is with the Research Center for Information Technology Innovation,
Academia Sinica, Taipei, Taiwan.}

\thanks{This work was supported in part by the
Ministry of Science and Technology under grants
100-2628-E-007-025-MY3 and 102-2221-E-007-016-MY3.}}



\maketitle


\vspace{-1cm}

{\renewcommand{\baselinestretch}{1.3}
\begin{abstract}
Physical layer multicasting with opportunistic user selection (OUS) is
examined in this work for multicell multi-antenna wireless systems.
In multicast applications, a common message is to be sent by the
base stations
to all users in a multicast
group.
By adopting a two-layer encoding scheme,
a rate-adaptive channel code is applied in each fading block to enable successful decoding by a chosen subset of users (which varies over different blocks) and an application layer erasure code is employed across multiple blocks to ensure that every user is able to recover the message after decoding successfully in a sufficient number of blocks.
The transmit signal and code-rate in each block determine opportunistically the subset of users that are able to successfully decode and can be chosen to maximize the long-term multicast efficiency.
The employment of OUS not only helps avoid rate-limitations caused by the user with the worst channel, but also helps coordinate interference among different cells and multicast groups.
In this work, efficient algorithms are proposed for the design of the transmit covariance matrices, the physical layer code-rates, and the target user subsets  in each block.
In the single group scenario, the system parameters are determined by maximizing the group-rate, which is defined  as the physical layer code-rate times the fraction of users that can successfully decode in each block.
In the multi-group scenario, the
system parameters are determined by considering a group-rate balancing optimization problem (i.e., a max-min weighted group-rate problem), which is solved using a
successive convex approximation (SCA) approach.
To further reduce the feedback overhead, we also consider the case where
only part of the users feed back their channel vectors in each block and
propose a design based on the balancing of the expected group-rates. In addition to applying SCA, a sample average approximation
technique is also introduced to handle the probabilistic terms that arise in this problem.
The effectiveness of the proposed schemes is demonstrated through computer simulations.
\end{abstract}
}


\section{Introduction}

\subsection{Motivation and Background}

Multicasting has attracted much attention in recent years
due to the increasing demand for mass content distribution, such as
software and firmware updates, file downloads, and multimedia streaming. In these applications, common information is to be disseminated efficiently to all users in a multicast group. Due to its importance in mobile applications, such service is also being introduced into current and next-generation cellular standards, such as the global system for mobile communications (GSM), the worldwide interoperability for microwave access (WiMAX), the long term evolution (LTE) etc., in the form of the so-called multimedia broadcast/multicast service (MBMS) \cite{gruber_zeller_2011,lecompte_gabin_2012,jiang_xiang_chen_ni_2007}. Different from wireline networks, multicasting in wireless systems enjoys the so-called wireless broadcast advantage (namely, the advantage that all users within the transmission range can receive) and, therefore, many research efforts have been devoted to the development of physical layer techniques that can fully exploit these advantages.

Most works in the literature on wireless physical layer multicasting consider the transmit signal design at the base station (BS) or BSs
for efficient delivery of common information to all users in a multicast group. Due to fading, the channels experienced by users in the multicast group may vary drastically and, thus, the rate of the channel code  must be low enough to ensure that all users in the group can successfully decode.
When multiple antennas are available at the transmitter, beamforming and precoding techniques can be further employed to improve the effective channel quality
of the worst user
\cite{sidiropoulos_davidson_luo_2006,lozano_2007,zhu_prasad_rangarajan_2012} and, thus, the multicast rate. With no restrictions on the rank of the signal covariance matrix, the optimal precoder (or the transmit covariance matrix) can be found using semi-definite programming (SDP) techniques \cite{jindal_luo_2006}. However, to reduce decoding complexity at the receivers, many works, such as \cite{sidiropoulos_davidson_luo_2006} and \cite{lozano_2007}, considered instead the use of multicast beamforming (i.e., signals with rank-$1$ covariance matrices) and proposed approximate algorithms for the design of multicast beamformers, which is otherwise known to be NP-hard \cite{sidiropoulos_davidson_luo_2006}. These works were extended to systems with multiple multicast groups in \cite{karipidis_sidiropoulos_luo_2008,silva_klein_2009,bornhorst_pesavento_2011,Christopoulos_Chatzinotas_Ottersten_2014TSP,Christopoulos_Chatzinotas_Ottersten_2014GLOBECOM}, where cochannel interference between different groups' messages was taken into consideration, and were also extended to multicell systems in \cite{xiang_tao_wang_2013,dartmann_gong_ascheid_2011}, where coordinated transmissions among BSs were employed. However, even with the above signal designs, the efficiency of the physical-layer multicast transmission is still fundamentally limited by the user
with the worst channel and the rate required for all users to decode may eventually go to zero as the number of users in the
group increases.

\vspace{-.2cm}

\subsection{Related Works on Opportunistic Multicasting}

\begin{figure}[t]
     \centering
     {\includegraphics[scale=1.2]{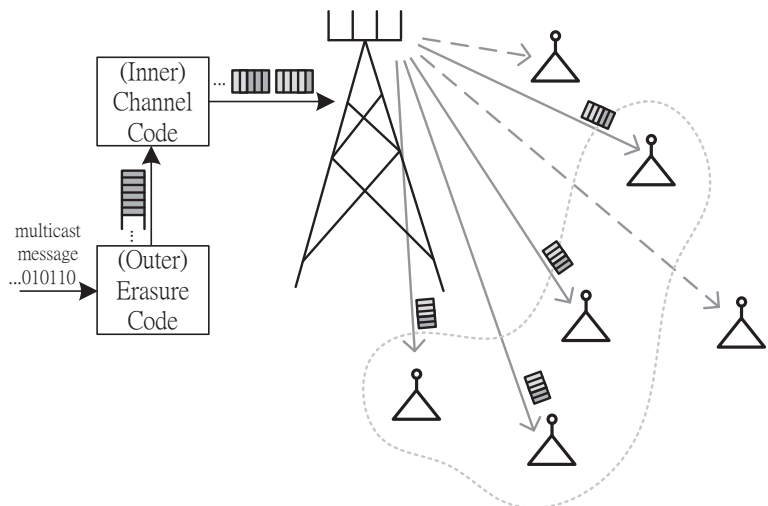}}
     \caption{Illustration of the opportunistic multicast scheduling (OMS) scheme.}
     \label{fig:OMSfigure}
\end{figure}

For delay-tolerant applications, such as file downloads and
software or firmware updates, the limitations caused by the worst
user in the network can be overcome by dividing the transmission
into multiple blocks and by scheduling only a subset of users to
receive in each block. This technique was referred to as \emph{opportunistic
multicast scheduling (OMS)} in \cite{kozat_2008,gopala_elgamal_2004,
gopala_elgamal_2005, low_pun_hong_kuo_2010}. In particular, the OMS scheme requires a
two-layer encoding scheme, as illustrated in Fig. \ref{fig:OMSfigure}, which consists of a physical layer channel code
(hereby referred to as the inner code) whose rate is adapted to the
chosen user subset in each block and an application layer erasure
code (referred to as the outer code) that is performed across
multiple blocks. By selecting only a subset of users to serve in each block, the rate
of the inner code would be less restrictive since successful
decoding needs to be guaranteed for fewer users in each block. The overall {\it group-rate}, defined as the code-rate
multiplied by the fraction of users in the group that
successfully decode, can thus be effectively increased. However, since a
user may not be able to successfully decode in every block, the
channel that it experiences across multiple blocks is effectively an
erasure channel, and an outer erasure code should be employed
to ensure that all users in the multicast group can eventually recover the message after decoding over a sufficient number of blocks. {The OMS scheme can achieve significant performance gains, but comes at the expense of delay. That is, these gains result from using time
as an additional resource or degree of freedom as compared to conventional
schemes that require successful and instantaneous decoding in every block.}

In practice, the outer code can be
implemented using, e.g., LT, Raptor, and Fountain codes
\cite{luby_etal_2007,mackay_2005}. With OMS, the group-rate is known
to converge to a non-zero constant as the number of users inceases
\cite{kozat_2008,gopala_elgamal_2004, gopala_elgamal_2005}. This is in contrast to cases without OMS where the group-rate is known to diminish to zero.
Most
works in the literature on OMS focused on the single-cell
single-antenna scenario, as in
\cite{kozat_2008,gopala_elgamal_2004, gopala_elgamal_2005,
low_pun_hong_kuo_2010}. Our work focuses instead on the multi-antenna scenario, where the problem is considerably more difficult due to the dependence between the user selection and the transmit signal design. The multi-antenna scenario was also examined more
recently in
\cite{low_etal_2010,kaliszan_pollakis_stanczak_2011,kaliszan_pollakis_stanczak_2012}, but only for single-cell scenarios. In particular, in \cite{low_etal_2010}, the joint precoding and user selection was performed using a heuristic semi-orthogonal vector selection algorithm.
Their proposed scheme has low complexity but is suitable only for single-cell scenarios with sum power constraints. In \cite{kaliszan_pollakis_stanczak_2011,kaliszan_pollakis_stanczak_2012},
the transmit signal design was restricted to beamforming and the
user selection was performed using a heuristic subset search
algorithm. The algorithm can be extended to multicell scenarios,
but is subject to high computational complexity.

\vspace{-.2cm}

\subsection{Main Contributions}

The main contribution of this work is the development of efficient algorithms for the joint design
of the transmit covariance matrix and the opportunistic user
selection policy in multicell networks with multiple multicast
groups. Here, group-rate (which is defined as
the physical layer code-rate in a certain block multiplied by the
fraction of users that can successfully decode) is utilized as the optimizing criterion. This is different from most works in the literature on physical layer multicasting, e.g., \cite{sidiropoulos_davidson_luo_2006,karipidis_sidiropoulos_luo_2008}, which typically do not consider user selection and, thus, can utilize the signal-to-interference-plus-noise ratio (SINR) as the optimizing criterion. In fact, the SINR criterion is not applicable when user selection is considered since it does not reflect the effect of the number of users served in each block. For example, if the system is optimized by maximizing the worst SINR among all selected users, then only one user (i.e., the user with the best effective channel) in each group would be selected. However, finding the optimal transmit covariance matrix and the opportunistic user selection (OUS) policy to maximize the group-rate may be difficult and is, in fact,
claimed to be NP-hard in
\cite{kaliszan_pollakis_stanczak_2011}.
Therefore, we propose in this work an approximate solution based on the introduction and relaxation of a set of
binary user selection variables. This technique is similar to that previously adopted in
\cite{matskani_sidiropoulos_luo_tassiulas_2008} and
\cite{matskani_sidiropoulos_luo_tassiulas_2009} for admission control problems.

In this work, we consider both single-group and multi-group
multicasting scenarios. We show that OUS can be especially effective in the latter case, where it not only can help avoid limitations by the
worst user but also can help coordinate interference among different
cells and multicast groups. In
the single group scenario, the optimization of the transmit covariance matrix and the user subset selection is performed first by relaxing the integer constraints corresponding to the
user selection variables and then by performing a sequential deflation technique, where users are eliminated one-by-one from an initial set of all users to yield candidate user subsets. In the multi-group scenario, the design
parameters are determined based on the group-rate balancing
criterion (also referred to as the max-min weighted group-rate
criterion) and the optimization problem is solved using a similar relaxation technique along with a
successive convex approximation (SCA) approach. In the above, the channel
state information (CSI) of all users is first assumed to be available at
the transmitter, which may be costly in practice. To reduce the
transmission overhead, we also consider the case where only a
subset of users feeds back their CSI in each block and
propose a design based on the
balancing of the expected group-rates. In addition to the SCA approach mentioned above, a sample average approximation (SAA) technique \cite{wang_ahmed_2008,pagnoncelli_ahmed_shapiro_2009} is further introduced to handle the probabilistic terms that may arise. The effectiveness of the
proposed schemes, compared to \cite{low_etal_2010,kaliszan_pollakis_stanczak_2011,kaliszan_pollakis_stanczak_2012}, is demonstrated through computer simulations.

In multicell systems, three cases are often considered based on different levels of BS cooperation \cite{gesbert_etal_2010,bjornson_etal_2010,dahrouj_yu_2010}, namely, full BS cooperation, interference coordination, and no BS cooperation. {\it Full BS cooperation} refers to the case where all BSs have knowledge of the information intended to all multicast groups and transmit cooperatively as a networked multiple-input multiple-output (MIMO) system. {\it Interference coordination} refers to the case where each BS serves only one multicast group and only has knowledge of the message intended for that group. In this case, cooperative transmission of common data is not possible, but the transmit signals can be designed to reduce cochannel interference among different multicast groups. {\it No BS cooperation} refers to the case where no CSI from users in other cells is available and thus no cooperation is adopted. Our system model is general enough to include all the cases mentioned above.
Moreover, we would also like to mention that, in this work, we do not restrict ourselves to rank-$1$ transmissions, as done in \cite{sidiropoulos_davidson_luo_2006,lozano_2007,karipidis_sidiropoulos_luo_2008,silva_klein_2009,bornhorst_pesavento_2011,kaliszan_pollakis_stanczak_2011,kaliszan_pollakis_stanczak_2012}, in order to maintain optimality \cite{jindal_luo_2006} and also to avoid distracting the readers, since a rank-$1$ constraint on the transmit covariance matrix is already enough to make the problem intractable \cite{sidiropoulos_davidson_luo_2006}, even without considering user selection. However, if rank-$1$ transmissions are desired, our solution can be used as the semi-definite relaxation (SDR) of the optimal beamformer design and the desired beamforming vectors can be extracted from our solutions using rank-$1$ approximation techniques, e.g., the Gaussian randomization procedure, as discussed in \cite{luo_etal_2010}.

The remainder of the paper is organized as follows. In Section \ref{sec.system_model}, we introduce the general multicell multicast scenario and the proposed group-rate balancing problem. In Sections \ref{sec.single_group} and \ref{sec.multiple_group}, the joint design of the transmit covariance matrix and the user selection policy is examined for the single-group and the multi-group scenarios, respectively. In Section \ref{sec.feedback}, the problem is extended to the case where only a subset of users feeds back their CSI in each block. Finally, computer simulations are provided in Section \ref{sec.simulations} and the paper is concluded in Section \ref{sec.conclusion}.

\section{System Model and Problem Formulation}\label{sec.system_model}


\begin{figure}[t]
     \centering
     {\includegraphics[scale=.8]{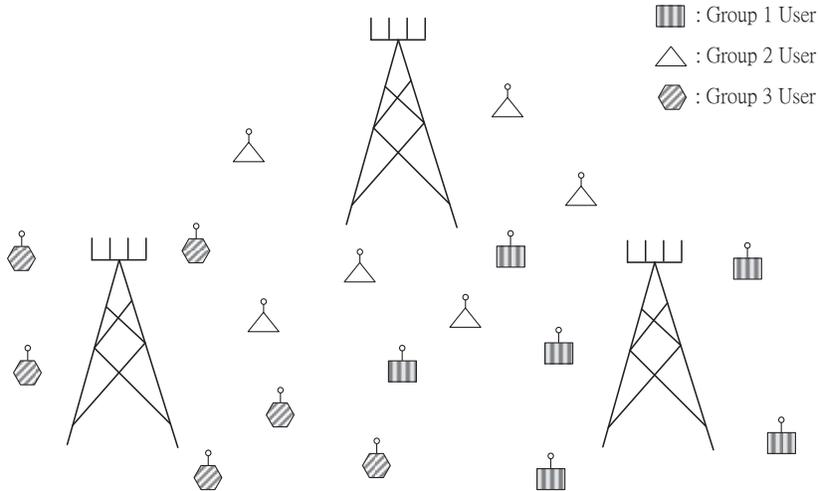}}
     \vspace{-.25cm}
     \caption{Illustration of a multicell network with $3$ multicast groups.}
     \vspace{-.45cm}
     \label{fig:system_model}
\end{figure}

Let us consider a downlink multicell network with $B$ BSs, each
equipped with $M$ antennas, and $K$ single-antenna users.
An illustration of the system under consideration is given in Fig.\,\ref{fig:system_model}.
The set of users $\calK$, with
cardinality $|\calK|=K$, is divided into $G$ multicast groups,
denoted by $\calK_1$, $\ldots$, $\calK_{G}$, where users in the same
group are interested in receiving the same multicast message.
Moreover, let ${\calB_g\subseteq\{1,\ldots, B\}}$ be the set of
BSs that are serving group $\calK_g$ and let
${\calG_b\subseteq\{1,\ldots, G\}}$ be the set of multicast
groups served by BS $b$. Notice that the case of full BS cooperation
can be considered by setting $\calB_g=\{1,\ldots,B\}$, for all $g$,
and $\calG_b=\{1,\ldots, G\}$, for all $b$, whereas the case of
interference coordination, where each BS has information intended
for only one multicast group, can be considered by setting $B=G$,
$\calB_g=\{g\}$ and $\calG_b=\{b\}$, for all $g$ and $b$. The case of no cooperation can also be considered by setting
$B=G=1$, $\calB_1=\{1\}$ and $\calG_1=\{1\}$, where the BS focuses on
sending information to the intended users without knowledge of other BSs' channels or transmitted signals (i.e., while neglecting the presence of other BSs).



Here, we consider a delay-tolerant application where  multicast
messages can be encoded over multiple coherence intervals (hereafter
referred to as \emph{blocks}). Let $\bs_{b,g}[n]$ be the signal
transmitted by BS $b\in\calB_g$ to users in group $g$ in block $n$. In this case, the received
signal at user $k\in\calK_g$ can be written as
\begin{align}
y_k[n]=\sum_{b \in \calB_g}\bh_{b,k}[n]^{H}\bs_{b,g}[n]+\sum_{b \in\calB_g}\bh_{b,k}[n]^{H} \sum_{\substack{\ell \in \calG_{b}\\\ell \neq g}}\bs_{b,\ell}[n]+\sum_{b' \notin \calB_g}{\bh_{b',k}[n]^{H}}\sum_{\ell \in \calG_{b'}}\bs_{b',\ell}[n]+n_k[n],\label{eq.signal1}
\end{align}
where $\bh_{b,k}[n]$ is the $M\times 1$ complex channel vector between BS $b$ and user $k$ in block $n$ and $n_k[n]$ is the complex additive white Gaussian noise (AWGN) at user $k$ with zero mean and unit variance, i.e., $n_k \sim \calC\calN(0,1)$. The entries of $\bh_{b,k}$ are assumed to be independent complex Gaussian random variables with variances that may be different for different $b$ and $k$, due to path loss or other large-scale fading effects. The first term in \eqref{eq.signal1} represents the sum of signals intended for user $k$, the second term represents the interference caused by the signals transmitted by user $k$'s serving BSs to users in other multicast groups, and the third term represents the interference caused by non-serving BSs.

By defining $\bh_k[n]=[\bh_{1,k}[n]^H,\ldots,\bh_{B,k}[n]^H]^H$ and $\bs_g=[\bs_{1,g}[n]^H,\ldots,\bs_{B,g}[n]^H]^H$, where $\bs_{b,g}={\bf 0}$ for $b\notin\calB_g$, the received signal at user $k$ can be rewritten as
\begin{equation}\label{eq.signal2}
y_k[n]=\bh_k[n]^H\bs_g[n]+\bh_k[n]^H\sum_{\ell \neq g}\bs_\ell[n]+n_k[n].
\end{equation}
%
%
%
%
Therefore, the SINR
at user $k \in \calG_g$ in block $n$ is given by
\begin{align}
\sinr_k[n]=\frac{\E[|\bh_k[n]^H\bs_g[n]|^2]}{\sum_{\ell
\neq
g}\E[|\bh_k[n]^H\bs_\ell[n]|^2]+1}=\frac{\text{tr}(\bQ_g[n]\bh_k[n]\bh_k[n]^H)}{\sum_{\ell
\neq g}\text{tr}(\bQ_\ell[n]\bh_k[n]\bh_k[n]^H)+1}
\end{align}
where $\bQ_g[n]=\E[\bs_g[n]\bs_g[n]^H]$.
Note that $\bQ_g[n]$ is an $MB\times MB$ matrix whose $(b,b')$-th block entry of dimension $M\times M$
is $\{\bQ_g[n]\}_{b,b'}\triangleq \E[\bs_{b,g}[n]\bs_{b',g}[n]^H]$. This block entry is zero if either $b$ or $b'$ does not belong to $\calB_g$.
The signals transmitted by BS $b$ must satisfy the per BS power constraint
\begin{equation}\label{eq.power_constraint}
\sum_{g\in\calG_b}\E[\|\bs_{b,g}[n]\|^2]=\sum_{g\in\calG_b}\tr(\{\bQ_g[n]\}_{b,b})\leq P_b.
\end{equation}

By employing OUS, only a subset of users in each group, i.e.,
${\calA_g[n]\subseteq\calK_g}$, for group $g$, is required to
decode successfully in block $n$. In this case, the rate of the
physical layer encoding (i.e., the inner code) of group $g$, i.e.,
$R_g[n]$, only needs to be low enough to ensure successful decoding
by users in $\calA_g[n]$. More specifically, by assuming that the
duration of each block is large enough to invoke Shannon's random
coding argument \cite{cover_thomas_2006}, the rate $R_g[n]$ of the
message intended for the subset of users $\calA_g[n]$ in block $n$
must be chosen such that
\begin{align}
R_g[n]\leq \log \left(1+
\sinr_k[n]\right)=\!\log\!\bigg(\!1\!+\!\frac{\text{tr}(\bQ_g[n]\bh_k[n]\bh_k[n]^H)}{\sum_{\ell
\neq g}\text{tr}(\bQ_\ell[n]\bh_k[n]\bh_k[n]^H)\!+\!1}\!\bigg) ~{\rm(bps/Hz)}\label{eq.rate_constraint}
\end{align}
for all $k\in\calA_g[n]$.

Since the codeword in each block is decoded only by a subset of users, each user effectively experiences an erasure channel across multiple blocks and, thus,  an outer (erasure) code, such as LT, Raptor, and Fountain codes \cite{luby_etal_2007,mackay_2005}, is needed to ensure eventual recovery of the original multicast data at all users. By assuming that an ideal erasure code is employed so that perfect erasure correction is performed, the average rate of user $k\in\calK_g$ over $T$ blocks can be written as
\begin{equation}
\frac{1}{T}\sum_{n=1}^T R_g[n]{\bf 1}_{\{k\in\calA_g[n]\}},
\end{equation}
where ${\bf 1}_{\{\cdot\}}$ is the indicator function, and the achievable multicast rate of group $g$ is thus given by
\begin{equation}\label{eq.min_gropu_rate}
\bar{R}_{g,\min}=\min_{k\in\calK_g}\frac{1}{T}\sum_{n=1}^TR_g[n]\mathbf{1}_{\{k\in\calA_g[n]\}}.
\end{equation}

Conventional physical layer multicast problems often use $\bar{R}_{g,\min}$ as the maximization criterion for system design.
However, maximizing $\bar{R}_{g,\min}$ would require non-causal knowledge of the channel realizations in the future $T$ fading blocks, which is not attainable in practice. In order to implement the user selection and transmit signal design in real-time, we propose to maximize the (average) group-rate defined as
\begin{equation}\label{eq.avg_group_rate}
\bar R_g\!=\!\frac{1}{|\calK_g|}\!\sum_{k\in\calK_g}\!\frac{1}{T}\sum_{n=1}^T R_g[n]{\bf 1}_{\{k\in\calA_g[n]\}}\!=\!\frac{1}{T}\!\sum_{n=1}^T\! \bar R_g[n],
\end{equation}
for group $g$, where
\begin{equation}\label{eq.inst_group_rate}
\bar R_g[n]=\frac{1}{|\calK_g|}R_g[n]\sum_{k\in\calK_g}{\bf 1}_{\{k\in\calA_g[n]\}}
\end{equation}
is the instantaneous group-rate in block $n$. The average group-rate provides a measure of the
average speed for which the users are able to acquire the multicast message. One can see that the maximization of the average group-rate can be achieved by maximizing the instantaneous group-rate block-by-block, and hence can be implemented in real-time. However, it should be noted that maximizing the average group-rate $\bar R_g$
may not be equivalent to maximizing the actual multicast rate $\bar{R}_{g,\min}$ in the general case. Therefore, $\bar R_g$ should only be viewed as an {\it auxiliary} maximization criterion that is used to allow for real-time optimization of the system parameters.
However, it was shown in \cite{kaliszan_pollakis_stanczak_2011} (and observed similarly for single antenna scenarios in \cite{kozat_2008} and \cite{low_pun_hong_kuo_2010}) that, when the channel vectors are independent and identically distributed (i.i.d.) across users and over time, the multicast rate $\bar{R}_{g,\min}$ is asymptotically equal to the average group-rate
$\bar R_g$ and, thus, is maximized asymptotically by maximizing the instantaneous group-rate $\bar R_g[n]$ in each block. The i.i.d. assumption is reasonable in many cases and is often considered when prior knowledge of the channel distributions or the users' locations are unavailable. When the channel vectors are non-i.i.d., normalization of the channel coefficients can be performed to ensure long-term fairness, as to be discussed in Sections \ref{sec.subsec.SGfair} and \ref{sec.subsec.MGfair}.


Maximizing the instantaneous group-rate for block $n$ involves optimizing over the transmit covariance matrices $\{\bQ_g[n]\}_{g=1}^G$, the user subsets $\{\calA_g[n]\}_{g=1}^G$, and the physical layer code-rates $\{R_g[n]\}_{g=1}^G$ subject to the power and rate constraints given in \eqref{eq.power_constraint} and \eqref{eq.rate_constraint}. This problem is examined for both the single-group and the multi-group multicasting scenarios. Since the optimization is performed separately in each block, we shall omit the block index $n$ in the remainder of this work.

In the single-group multicasting scenario (i.e., the case where $G=1$), the optimal system parameters can be determined by solving the following instantaneous group-rate maximization (GRM) problem.

\noindent\underline{\bf Group-Rate Maximization (GRM) Problem:}
\begin{subequations}\label{eq.GRM1}
\begin{align}
\text{maximize} \quad & \frac{1}{|\calK|}R\sum_{k\in\calK}{\bf 1}_{\{k\in\calA\}} \\
\text{subject to}\quad & \log_2[ 1+ \text{tr}(\bQ\mathbf{h}_{k}\mathbf{h}_{k}^{H})] \geq R, ~\text{ for }k\in\calA,\label{eq.GRM1.const1}\\
& \text{tr}(\{\bQ\}_{b,b}) \leq P_{b}, ~\forall b,~~ \mathbf{Q} \succeq \mathbf{0}.\\
\text{variables:}\quad & \bQ,\calA, R.\notag
\end{align}
\end{subequations}
Notice that no cochannel interference exists in this case and that the group index $g$ is omitted since only a single multicast group is considered.  This problem is claimed to be NP-hard in \cite{kaliszan_pollakis_stanczak_2011} and no known solutions are available to solve this problem efficiently. An approximate solution will be studied in Section \ref{sec.single_group} based on the introduction of a set of binary user selection variables and a convex relaxation of the problem.

In the multigroup multicasting scenario (i.e., the case where $G>1$), we consider the following group-rate balancing (GRB) optimization problem where the system parameters are jointly determined to maximize the worst weighted group-rate among all multicast groups.


\noindent\underline{\bf Group-Rate Balancing (GRB) Problem:}
\begin{subequations}\label{eq.GRB1}
\begin{align}
\text{maximize}~~ & \mathop{\min_{  g \in \{ 1,...,G\}  }}\frac{1}{\tau_g}\frac{1}{|\calK_g|} \,R_g\!\sum_{k\in\calK_g}{\bf 1}_{\{k\in\calA_g\}} \\
\text{subject to}~~ &  \log_{2}\left( 1+ \frac{\text{tr}(\bQ_g\bh_k\bh_k^H)}{\sum_{\ell \neq g}{\text{tr}(\bQ_{\ell}\bh_k\bh_k^H)+1}}  \right) \geq R_g, \forall k \in \calA_g, ~\forall g, \\
&
 \sum_{g=1}^G{\text{tr}(\{\bQ_g\}_{b,b})} \leq P_{b}, ~\forall b,~~  \bQ_g \succeq \mathbf{0}, ~\forall g,\\
& \{\bQ_g\}_{b,b'}\!=\!{\bf 0}_{M\!\times\! M}, \text{ for } b\notin\calB_g
\text{ or }b'\notin\calB_g,\\
\text{variables:}~~ & \{\bQ_g\}_{g=1}^G,\{\calA_g\}_{g=1}^{G},\{R_g\}_{g=1}^{G},\notag
\end{align}
\end{subequations}
where $\tau_g$ is the parameter that specifies the priority of group
$\calK_g$. {Solving the above problem guarantees a common level of quality-of-service (QoS) for all multicast groups. In particular, if the
resulting objective value is $\alpha$, then the group-rate of
group $g$ will be at least $\tau_g\alpha$, for all $g$.} It is worthwhile to note
that the GRB problem is analogous to the SINR-balancing problem
often considered for the design of multiuser downlink beamforming
\cite{schubert_boche_2004} or multigroup multicast beamforming
\cite{karipidis_sidiropoulos_luo_2008} schemes. While the SINR
criterion may be suitable for conventional multicast beamforming
designs, where all users are required to decode successfully in each
fading block, it is not suitable for systems employing OUS. This is
because, when considering the SINR-balancing (or, equivalently, the
max-min SINR) criterion, the optimization will result in a trivial
user selection policy where only the user with the best channel in
each group is chosen. Even though the physical layer code-rate can
be chosen to be the highest in this case, the group-rate is limited
by the fact that only one user is able to decode. The GRB
formulation is interesting in the sense that it allows us to take
into consideration the effect of user selection as well as  to
incorporate the impact of QoS requirements,
i.e., $\tau_g$, into the problem. This problem will be examined
further in Section \ref{sec.multiple_group} and an approximate
solution will be proposed based on a relaxation similar to that in the
single-group scenario.

\begin{remark}\it
Most works in the literature on physical layer multicasting (e.g., \cite{sidiropoulos_davidson_luo_2006,lozano_2007,zhu_prasad_rangarajan_2012,karipidis_sidiropoulos_luo_2008,silva_klein_2009,bornhorst_pesavento_2011,xiang_tao_wang_2013,dartmann_gong_ascheid_2011}) assume that all users must decode successfully in each block.
In this case, the outer erasure code is not needed and the instantaneous group-rate will be equal to the physical layer code-rate in each block. The group-rate in block $n$ must then satisfy
\[\bar R_g[n]=R_g[n]\leq \log \left(1+ \sinr_k[n]\right), ~~\forall
k\in\calK_g.
\]
The conventional approach can be viewed as the case where we choose $\calA_g[n]=\calK_g$, for all $g$ and $n$. The proposed OUS scheme instead optimizes over $\calA_g[n]$ and, thus, the achievable group-rate can be no less than that in conventional systems.
\end{remark}


\section{Group-Rate Maximization for Single-Group Multicasting}\label{sec.single_group}

In this section, we examine the GRM problem in \eqref{eq.GRM1} for the single-group multicasting scenario.
Note that the joint design of all system parameters in this problem is in general NP-hard \cite{kaliszan_pollakis_stanczak_2011} due to the combinatorial nature of the user subset selection. However, as similarly observed in \cite{kaliszan_pollakis_stanczak_2011}, the problem reduces to a standard SDP problem when the user subset is given. In particular, for a given user subset $\calA$, the SDP problem can be formulated as follows:
\begin{subequations}\label{eq.GRMwithA1}
\begin{align}
\max\quad &  R\cdot |\calA|  \\
\text{subject to}\quad & \log_2[ 1+ \text{tr}(\bQ\mathbf{h}_{k}\mathbf{h}_{k}^{H})] \geq R, ~\text{ for }k\in\calA,\\
& \text{tr}(\{\bQ\}_{b,b}) \leq P_{b}, ~\forall b,  ~~\mathbf{Q} \succeq \mathbf{0},\\
\text{variables:}\quad & \bQ, R.\notag
\end{align}
\end{subequations}
Due to this observation, the authors in  \cite{kaliszan_pollakis_stanczak_2011} proposed a heuristic subset search algorithm where all users are considered initially and one user is excluded from the set in each iteration until no further gain in the objective value is obtained. The user removed in each iteration is the user whose removal from the set results in the maximum increase in the objective value. Let $g(\calA)$ be the optimal objective value of \eqref{eq.GRMwithA1} for a given user subset $\calA$. Then, the subset search algorithm can be summarized as follows.

\begin{algorithm}[h!]
\caption{Subset Search Algorithm \cite{kaliszan_pollakis_stanczak_2011}}\label{alg.subset}
    \begin{enumerate}
        \item Set $\calA=\{1,\ldots, K\}$.
        \item Define $\calA_{-k}\triangleq\calA-\{k\}$, for all $k\in\calA$, and let $k^*=\arg\max_k g(\calA_{-k})$.
        \item If $g(\calA)>g(\calA_{-k^*})$, then stop and take $\calA$ as the solution; else, set $\calA\leftarrow\calA-\{k^*\}$ and go to Step 2.
    \end{enumerate}
\end{algorithm}

Notice from Algorithm \ref{alg.subset} that, in evaluating $k^*$ in each iteration, the SDP problem in \eqref{eq.GRMwithA1} needs to be solved for every possible choice of $\calA_{-k}$. This requires one to solve the above SDP problem in the order of $O(K^2)$ times, which can be inefficient as $K$ increases.

\subsection{Reformulation and Relaxation of the GRM Problem using User Selection Variables}

To solve the GRM problem in \eqref{eq.GRM1} more efficiently, let us first introduce a set of binary user selection variables $\{s_k, \forall k\in\calK\}$, where $s_k=1$ if user $k$ is selected (i.e., if $k\in\calA$) and $s_k=0$, otherwise.
By choosing $\delta$ to be sufficiently small, the GRM problem can be equivalently formulated as
\begin{subequations}\label{eq.GRM2}
\begin{align}
\max \quad & \frac{1}{|\calK|} R \sum_{k\in\calK}s_k \\
\text{subject to}\quad & \log_2[ 1+ \text{tr}(\bQ\mathbf{h}_{k}\mathbf{h}_{k}^{H})] +\delta^{-1}(1-s_k)\geq R, ~\forall k\in\calK,\label{eq.GRM2.const1}\\
& \text{tr}(\{\bQ\}_{b,b}) \leq P_{b}, ~\forall b, ~~\mathbf{Q} \succeq \mathbf{0},\\
&
s_k\in\{0,1\}, ~\forall k,\label{eq.GRM2.const3}\\
\text{variables:}\quad & \bQ, R, \{s_k\}_{k\in\calK}.\notag
\end{align}
\end{subequations}
The equivalence of the two problems is stated in the following  lemma. The proof is given in Appendix \ref{app.lemma}.
\vspace{.1cm}
\begin{lemma}\label{lemma.GRMequiv}\it
For $\delta\leq
\left[\max_k\log_2(1+\sum_{b=1}^BP_b\|\bh_{b,k}\|^2)\right]^{-1}$, $(\bQ^*,R^*,\{s_k^*\}_{k\in\calK})$ is an optimal solution of
the problem in \eqref{eq.GRM2} if and only if $(\bQ^*, R^*, \calA^*)$, where
$\calA^*\triangleq\{k\in\calK:s_k^*=1\}$, is an optimal solution of
the GRM problem in \eqref{eq.GRM1}.
\end{lemma}
\vspace{.1cm}

Notice that the problem in \eqref{eq.GRM2} is non-convex due to the integer constraints on $s_k$'s in \eqref{eq.GRM2.const3}. To obtain an efficient solution, we consider a relaxation of the problem  where the integer constraints on $s_k$'s are replaced with the linear constraints $0\leq s_k\leq 1$, for all $k$. By taking the logarithm of the objective and by omitting the irrelevant variables, the relaxed problem can be written as
\begin{subequations}\label{eq.GRMrelax1}
\begin{align}
\max \quad & \ln R +\ln \sum_{k\in\calK}s_k \\
\text{subject to}\quad & \log_2[ 1+ \text{tr}(\bQ\mathbf{h}_{k}\mathbf{h}_{k}^{H})] +\delta^{-1}(1-s_k)\geq R, ~\forall k\in\calK,\\
& \text{tr}(\{\bQ\}_{b,b}) \leq P_{b}, ~\forall b, ~~\mathbf{Q} \succeq \mathbf{0},\\
& 0\leq
s_k\leq 1, \forall k,\\
\text{variables:}\quad & \bQ, R, \{s_k\}_{k\in\calK}.\notag
\end{align}
\end{subequations}
This problem is convex and can be solved efficiently using general
purpose solvers such as CVX \cite{cvx}. By relaxing the integer
constraints on the user selection variables, the term
$\delta^{-1}(1-s_k)$ can take on any value between $0$ and
$\delta^{-1}$, and can be viewed as a measure of rate violation, i.e., the difference between the code-rate $R$ and the achievable rate of user $k$.
Similar approaches have also been used in studies of admission control
problems in \cite{matskani_sidiropoulos_luo_tassiulas_2008} and
\cite{matskani_sidiropoulos_luo_tassiulas_2009}. Notice that the
solution of $s_k$ in the relaxed problem may not take on the value $0$ or $1$. To convert the solution of the
relaxed problem to a feasible solution of \eqref{eq.GRM2}, we
propose a sequential deflation technique, where the $D$ users with
the $D$ smallest values of $s_k$ are removed in each iteration,
after which the values of $\bQ$, $R$, and  $s_k$, $\forall k$, are
updated by solving \eqref{eq.GRMrelax1} again. The value of $D$ can
be chosen as $1$ in most cases but can be chosen to be greater than
one for complexity reduction. Among the {$\lceil{K/D}\rceil$}
possible subsets obtained from this procedure, {where
$\lceil{a}\rceil$ is the smallest integer not less than
$a$}, the user subset that yields the largest group-rate is chosen.
%
%
The sequential deflation algorithm is summarized in Algorithm
\ref{alg.deflation}.

\begin{algorithm}[h!]
\caption{Opportunistic User Selection by Sequential Deflation}\label{alg.deflation}
    \begin{enumerate}
        \item[(i)] Initialize by setting $\calA\leftarrow\calK$, $\calA^*\leftarrow\emptyset$, and $\alpha^*\leftarrow0$.
        \item[(ii)] Solve \eqref{eq.GRMwithA1} for given user subset $\calA$ and let $\tilde \alpha$ be the resulting objective value.
        \item[(iii)] If $\tilde\alpha>\alpha^*$, then set $\alpha^*\leftarrow\tilde\alpha$ and
            $\calA^*\leftarrow\calA$.
        \item[(iv)] Solve the relaxed problem \eqref{eq.GRMrelax1} for $\calK=\calA$ to yield the values of $s_k$ for all $k\in\calA$.
        \item[(v)] Set $\calA\leftarrow \calA-\calS_{\min}$, where $\calS_{\min}$ is the set of users associated with the $D$ smallest values of $s_k$ among users in $\calA$.
 \item[(vi)] Repeat steps (ii)-(v) until $\calA=\emptyset$. Then, take $\calA^*$ as the desired user subset.
    \end{enumerate}
\end{algorithm}

Notice that the sequential deflation algorithm proposed above is
different from that proposed in
\cite{matskani_sidiropoulos_luo_tassiulas_2009} where the algorithm
is to terminate whenever the removal of a user no longer results in
an increase in group-rate. However, when applying their approach to
our problem, the algorithm often terminates early in the process at
a local optimum that is close to choosing all users as the serving subset.
Moreover, it is worthwhile to remark that, in Algorithm \ref{alg.deflation}, two convex optimization
problems are solved in each iteration (i.e., one for computing the relaxed
problem \eqref{eq.GRMrelax1} and one for solving
\eqref{eq.GRMwithA1} for given $\calA$) and the number of
iterations is equal to the number of users $K$ in the worst case, when $D=1$.
This is a significant improvement over the $O(K^2)$ worst-case
complexity required in the subset search method summarized in
Algorithm \ref{alg.subset} \cite{kaliszan_pollakis_stanczak_2011}.


\subsection{Fairness of the GRM Problem in the Non-I.I.D. Case based on Channel Normalization}\label{sec.subsec.SGfair}

Notice that maximizing the instantaneous group-rate, as done in the previous subsection, equivalently maximizes the average group-rate in \eqref{eq.avg_group_rate}. When the channel vectors (i.e., the short-term fading coefficients) are i.i.d. across users,  all users will have equal probability of being selected in each block and, thus, the average group-rate would eventually be the rate achieved by all users as $T\rightarrow \infty$ \cite{kaliszan_pollakis_stanczak_2011}. However, when the channel vectors are non-identically distributed, users with better average channels (namely, users closer to the BSs) may have a higher probability of being selected under our proposed scheme. In this case, an issue of fairness may arise, similar to that observed in \cite{kozat_2008} and \cite{low_pun_hong_kuo_2010} for the single-antenna case. In this section, a heuristic channel normalization technique is proposed to address the aforementioned fairness issue.

The key idea is to normalize the channel vectors of the users by their long term statistics and perform the proposed OUS based on the normalized channel vectors (which will then be i.i.d.). More specifically, suppose that the channel vector $\bh_{b,k}[n]$ between BS $b$ and user $k$ in block $n$ has entries that are i.i.d. $\calC\calN(0,d_{b,k}^{-\alpha})$, where $d_{b,k}$ is the distance between BS $b$ and user $k$ and $\alpha$ is the path loss exponent. Moreover, let $\bar d_b=\frac{1}{|\calK|}\sum_{k\in\calK} d_{b,k}$ be the average distance of all users to BS $b$. Let us define the normalized channel vector between BS $b$ and user $k$ in block $n$ as
\begin{equation}\label{eq.normalized_channel_SG}
\tilde \bh_{b,k}[n]=\bh_{b,k}[n]\sqrt{\frac{{\bar d_b}^{-\alpha}}{d_{b,k}^{-\alpha}}},
\end{equation}
which has entries that are i.i.d. $\calC\calN(0,{\bar d_b}^{-\alpha})$. The normalized channel vectors are then utilized to compute the optimal user subset, denoted by $\tilde \calA^*$, using the algorithm proposed in the previous subsection. The optimal transmit covariance matrix can then be computed by solving \eqref{eq.GRMwithA1} for given $\tilde\calA^*$ using the original channel vectors $\bh_{b,k}$, $\forall b,k$.

The normalized channel vectors are scaled by their average distance so that the distance between the BSs and the users are
taken into account in the signal-to-noise ratio (SNR) at the receiver. After normalization, the channel vectors experienced by all users will become i.i.d. and, by performing OUS based on the normalized channel vectors, all users will have equal probability of being selected in each block, ensuring fairness among users. The users that are opportunistically selected in each block will likely be users whose instantaneous channel gains are larger than their respective averages, exploiting the advantages of both temporal and multiuser diversity. It is also interesting to note that, even though the users in the subset $\tilde\calA^*$ are chosen as target users and the transmit covariance matrix $\bQ$ is chosen to maximize the rate of users in $\tilde\calA^*$, it is possible that users outside of the target subset $\tilde\calA^*$ may also be served due to their large average channel gains (i.e., their close distance to the BSs) or because of their coincidental locations in the directions of the beamformed signals. The performance of the proposed fair OUS scheme is demonstrated through simulations in Section \ref{sec.simulations}.

It is necessary to note that the proposed normalization scheme is applicable to any case in which the average channel gain is different for different users, not limited to that caused by path loss.
For frequency-selective fading scenarios, multicarrier modulation schemes such as OFDM can be considered and the proposed algorithm, including the channel normalization technique, can be applied to each subcarrier individually.


\section{Group-Rate Balancing for Multigroup Multicasting}\label{sec.multiple_group}

In this section, we examine in more detail the GRB problem described in \eqref{eq.GRB1} for the multigroup multicasting scenario. The problem is more challenging than that in the single-group scenario because of the existence of cochannel interference among different groups.

%

Notice that, similar to the single-group scenario, the complexity of solving the GRB problem in \eqref{eq.GRB1} is largely due to the combinatorial nature of the search for the optimal user subsets $\{\calA_g\}_{g=1}^G$. However, different from the problem in \eqref{eq.GRMwithA1}, the optimization problem does not reduce to a convex optimization problem even when the user subsets $\{\calA_g\}_{g=1}^G$ are given. Specifically, when $\{\calA_g\}_{g=1}^G$ are given, we have
\begin{subequations}\label{eq.GRBwithA1}
\begin{align}
\mathop{\max}~ & \mathop{\min_{  g\in \{ 1,...,G\}  }}\frac{1}{\tau_g'} R_g \\
\text{subject to}~ &  \log_{2}\bigg( 1+ \frac{\tr(\bQ_g\bh_k\bh_k^H)}{\sum_{\ell \neq g}\tr(\bQ_\ell\bh_k\bh_k^H)+1}  \bigg) \geq R_g, ~ \forall k \in \calA_g, \forall g, \label{eq.GRBwithA1const1}\\
& \sum_{g=1}^G\tr(\{\bQ_g\}_{b,b}) \leq P_{b}, ~\forall b,~~  \bQ_g \succeq \mathbf{0}, ~\forall g,\\
& \{\bQ_g\}_{b,b'}\!=\!{\bf 0}_{M\!\times\! M}, \text{ for } b\notin\calB_g
\text{ or }b'\notin\calB_g,\\
\text{variables:}~ & \{\bQ_g\}_{g=1}^G,\{R_g\}_{g=1}^G,\notag
\end{align}
\end{subequations}
where $\tau_g'\triangleq \tau_g|\calK_g|/|\calA_g|$. We can observe
that the rate variables $\{R_g\}_{g=1}^G$ are unconstrained
below and that the objective value does not become smaller by
decreasing $R_g/\tau_g'$ to the value
${\min_{g\in\{1,\dots,G\}}}R_g/\tau_g'$ for all
$g=1,\dots,G$. Consequently, we can impose the constraint
\[
R_1/\tau_1'=\cdots=R_G/\tau_G'
\]
on problem \eqref{eq.GRBwithA1} without loss of optimality.

By introducing the variable
$\alpha=R_1/\tau_1'=\cdots=R_G/\tau_G'$, the problem in
\eqref{eq.GRBwithA1} can be equivalently expressed as
\begin{subequations}\label{eq.GRBwithA2}
\begin{align}
\mathop{\max}~ & \alpha \\
\text{subject to}~&  \tr(\bQ_g\bh_k\bh_k^H) \geq \left( 2^{\tau_g'\alpha}-1\right)\left( \sum_{\ell \neq g}\tr(\bQ_\ell\bh_k\bh_k^H)+1\right),~ \forall k \in \calA_g, \forall g, \label{eq.GRBwithA2const1}\\
&
 \sum_{g=1}^G{\tr(\{\bQ_g\}_{b,b})} \leq P_{b}, ~\forall b,~~  \bQ_g \succeq \mathbf{0}, \forall g,\\
& \{\bQ_g\}_{b,b'}\!=\!{\bf 0}_{M\!\times \!M}, \text{ for } b\notin\calB_g
\text{ or }b'\notin\calB_g,\\
\text{variables:}~ & \{\bQ_g\}_{g=1}^G,~\alpha.\notag
\end{align}
\end{subequations}
This problem is still non-convex due to the constraint in
\eqref{eq.GRBwithA2const1}. However, for a fixed $\alpha$, the
constraint becomes convex and the problem becomes a convex
feasibility problem. The optimal value of $\alpha$ can then be found
via a bisection search as described in the following.

To perform the bisection search, it is necessary to first determine an upper bound and a lower bound on the value of $\alpha$. To do so, notice that, in \eqref{eq.GRBwithA2}, $\tau_g'\alpha=\tau_g\alpha|\calK_g|/|\calA_g|$ can be viewed as the minimum rate achievable by all users in the subset $\calA_g$ and, thus, is upper-bounded by the rate achievable when all BS powers are used to beamform data to the best user in group $g$ (and no signal is sent to other groups), i.e.,
\begin{equation}\label{eq.upperU_GRBwithA1}
\alpha\frac{\tau_g|\calK_g|}{|\calA_g|}\leq \max_{k\in\calA_g} \log_2\left(1+\sum_{b=1}^BP_b \|\bh_{b,k}\|^2\right),
\end{equation}
where $\bh_{b,k}$ is the channel from BS $b$ to user $k$. This inequality must hold for all $g$ and, thus, $\alpha$  is upper-bounded by
\begin{equation}\label{eq.upperU_GRBwithA2}
\alpha\!\leq\! \min_{g\in\{1,\ldots,
G\}}\max_{k\in\calA_g}\frac{|\calA_g|}{\tau_g|\calK_g|}
\log_2\left(\!1\!+\!\sum_{b=1}^BP_b \|\bh_{b,k}\|^2\!\right).
\end{equation}
Moreover, since the rate is non-negative, $\alpha$ is trivially lower-bounded by $0$.  The bisection algorithm repeatedly bisects the interval between the upper and lower bounds given above until the solution is obtained. Details are given in Algorithm \ref{alg.bisection}.


\begin{algorithm}[h!]
\caption{Bisection Search for $\alpha$}\label{alg.bisection}
\begin{enumerate}
\item Initialize $U$ as in \eqref{eq.upperU_GRBwithA2} and $L=0$.
\item Set $\alpha=\frac{U+L}{2}$ and solve the feasibility problem corresponding to \eqref{eq.GRBwithA2}. If it is feasible, then set $L=\alpha$. Otherwise, set $U=\alpha$.
\item Repeat Step 2 until $U-L \leq \epsilon$, where $\epsilon$ is the value specifying the convergence criteria. The resulting $\{\bQ_g\}_{g=1}^{G}$ and $\alpha$ are the desired solution to \eqref{eq.GRBwithA2}.
\end{enumerate}
\end{algorithm}

In the above, the optimal transmit covariance matrices $\{\bQ_g\}_{g=1}^{G}$ (and the corresponding rates $\{R_g\}_{g=1}^G$) were found for given user subsets $\{\calA_g\}_{g=1}^{G}$. To find $\{\calA_g\}_{g=1}^{G}$, one can employ a heuristic OUS policy similar to the subset search algorithm described in Algorithm\,\ref{alg.subset}. However, as mentioned previously, this requires solving \eqref{eq.GRBwithA2} in the order of $O(K^2)$ times, with a bisection search embedded in each computation. In the following, we propose a more efficient method based on the introduction of binary user selection variables, as done in the single-group scenario. Here, an SCA approach is further adopted to cope with the non-convexity caused by the interference terms.

\subsection{Reformulation and Relaxation of the GRB Problem using User Selection Variables}

Specifically, following the technique given in Section
\ref{sec.single_group}, let us introduce the set of binary user selection
variables $\{s_k, \forall \calK_g,\forall g\}$, where $s_k=1$ if $k\in\calA_g$ for some $g$, and
$s_k=0$, otherwise. Then, for a sufficiently small $\delta$, the GRB
problem in \eqref{eq.GRB1} can be equivalently formulated as
\begin{subequations}\label{eq.GRB2}
\begin{align}
\mathop{\max}~ & \mathop{\min_{  g \in \{ 1,...,G\}  }}\frac{1}{\tau_g|\calK_g|}\cdot R_g\sum_{k\in\calK_g}s_k \\
\text{subject to}~ & \log_2\left( \!1\!+\! \frac{\tr(\bQ_g\bh_k\bh_k^H)}{\sum_{\ell \neq k}\tr(\bQ_\ell\bh_k\bh_k^H)\!+\!1}  \right)+\delta^{-1}(1\!-\!s_k) \geq R_g, ~ \forall k \in\calK_g,~ \forall g,\\
&
 \sum_{g=1}^G{\tr(\{\bQ_g\}_{b,b})} \leq P_{b}, ~\forall b,~~  \bQ_g \succeq \mathbf{0}, \forall g,\label{eq.GRB2const1}\\
& \{\bQ_g\}_{b,b'}\!=\!{\bf 0}_{M\!\times \!M}, \text{ for } b\notin\calB_g \text{ or }b'\notin\calB_g,\label{eq.GRB2const2}\\
&  s_k\in\{0,1\}, ~\forall k\in\calK,\\
\text{variables:}~ & \{\bQ_g\}_{g=1}^G,\{R_g\}_{g=1}^{G},\{s_k\}_{k\in\calK}.\notag
\end{align}
\end{subequations}
The equivalence of the problems in \eqref{eq.GRB1} and \eqref{eq.GRB2} is stated in the following lemma.
\vspace{.1cm}
\begin{lemma}\it For
$$\delta\leq
\left[\max_{g\in\calG}\max_{k\in\calK_g}\log_2(1+\sum_{b=1}^BP_b\|\bh_{b,k}\|^2)\right]^{-1},$$
$(\{\bQ^*_g\}_{g=1}^G,\{R_g^*\}_{g=1}^G,\{s_k^*\}_{k\in\calK})$
is an optimal solution of the problem in \eqref{eq.GRB2} if and only if
$(\{\bQ^*_g\}_{g=1}^G,$ $\{R_g^*\}_{g=1}^G,\{\calA_g^*\}_{g=1}^G)$,
where $\calA^*_g\triangleq\{k\in\calK_g:s_k^*=1\}$, for $g=1,\ldots,
G$, is an optimal solution of the GRB problem in \eqref{eq.GRB1}.
\end{lemma}
\vspace{.1cm}

The proof is similar to that of Lemma \ref{lemma.GRMequiv} and, thus,  is
omitted. Moreover, let us also consider a relaxation where the integer
constraints on $s_k$'s are replaced with the linear constraints
$0\leq s_k\leq 1$, for all $k$. By introducing the auxiliary
variable $\alpha$, the relaxed problem can be written in the
epigraph form
\begin{subequations}\label{eq.GRBrelaxed1}
\begin{align}
\max~ & \alpha \\
\text{subject to}~ & R_g\sum_{k\in\calK_g}s_k\geq \tau_g|\calK_g|\alpha^2, ~\forall g,\label{eq.GRBrelaxed1const1}\\
& r_k(\{\bQ_\ell\}_{\ell=1}^G) +\delta^{-1}(1-s_k) \geq R_g,  \forall k \in\calK_g, \forall g,\label{eq.GRBrelaxed1const2}\\
& 0\leq s_k \leq 1, ~\forall k\in\calK,~\eqref{eq.GRB2const1},\text{
and } \eqref{eq.GRB2const2},\\
\text{variables:}~ & \{\bQ_g\}_{g=1}^G,\{R_g\}_{g=1}^{G},\{s_k\}_{k\in\calK}, \alpha,\notag
\end{align}
\end{subequations}
where
\begin{equation}
r_k(\{\bQ_\ell\}_{\ell=1}^G)\triangleq\log_2\left(1+\frac{\tr(\bQ_g\bh_k\bh_k^H)}{\sum_{\ell{\ne}g}\tr(\bQ_\ell\bh_k\bh_k^H)+1}\right),
\end{equation}
for $k\in\calK_g$. Notice that this problem cannot be solved using the bisection search algorithm
since, even when $\alpha$ is given, the problem does not reduce to a convex feasibility problem. This is because the value of $R_g$ cannot be
determined explicitly and the constraint in
\eqref{eq.GRBrelaxed1const2} is non-convex. To address this issue, we  employ an SCA approach as described in the
following.

First, notice that, since $R_g$ is positive, the non-convex constraint in \eqref{eq.GRBrelaxed1const1} can be written as
\begin{equation}
\sum_{k\in\calK_g}s_k-\alpha\sqrt{\tau_g|\calK_g|} R_g^{-1} \alpha\sqrt{\tau_g|\calK_g|}\geq 0.
\end{equation}
Then, by applying the Schur complement \cite{boyd_vandenberghe_2004}, this constraint can be equivalently written as the linear matrix inequality constraint
\begin{equation}\label{eq.GRBrelaxed1const1a}
\left(\begin{array}{cc}R_g & \alpha\sqrt{\tau_g|\calK_g|}\\
\alpha\sqrt{\tau_g|\calK_g|} & \sum_{k\in\calK_g}s_k
\end{array}\right)\succeq\mathbf{0}.
\end{equation}
Secondly, to address the non-convexity of the constraint in \eqref{eq.GRBrelaxed1const2},
we adopt an SCA technique similar to that employed in
\cite{li_chang_lin_chi_2013}. Specifically, given any
$\{\tilde{\bQ}_\ell\}_{\ell=1}^G$ satisfying \eqref{eq.GRB2const1}
and \eqref{eq.GRB2const2}, {which together with $\alpha=0$,
$R_g=\delta^{-1}$, $s_k=0$, $\forall{g,k}$, is a feasible point to
\eqref{eq.GRBrelaxed1},} we can rewrite $r_k(\{\bQ_\ell\}_{\ell=1}^G)$ and obtain its lower bound as
\begin{align*}
&r_k(\{\bQ_\ell\}_{\ell=1}^G)\\
&=\log_2\!\left(\!1\!+\!\sum_{\ell=1}^G\tr(\bQ_\ell\bh_k\bh_k^H)\!\right)\!-\!\log_2\left(\!1\!+\!\sum_{\ell{\ne}g}\tr(\bQ_\ell\bh_k\bh_k^H)\!\right)\\
&\ge\log_2\!\left(\!1+\!\sum_{\ell=1}^G\tr(\bQ_\ell\bh_k\bh_k^H)\!\right)\!-\!\log_2\left(\!1\!+\!\sum_{\ell{\ne}g}\tr(\tilde{\bQ}_\ell\bh_k\bh_k^H)\right)-\frac{\sum_{\ell{\ne}g}\tr((\bQ_\ell-\tilde{\bQ_\ell})\bh_k\bh_k^H)}{[1+\sum_{\ell{\ne}g}\tr(\tilde{\bQ}_\ell\bh_k\bh_k^H)]\ln 2}\\
&\triangleq\bar{r}_k(\{\bQ_\ell\}_{\ell=1}^G\mid\{\tilde{\bQ}_\ell\}_{\ell=1}^G)
\end{align*}
where the inequality comes from the first-order condition of the
concave function $\log_2(\cdot)$, i.e., $\log_2{y}\leq \log_2{x}+ \frac{1}{x\ln{2}}(y-x)$ for any $x,y>0$. Note that $\bar{r}_k(\{\bQ_\ell\}_{\ell=1}^G\mid\{\tilde{\bQ}_\ell\}_{\ell=1}^G)$ is concave in $\{\bQ_\ell\}_{\ell=1}^G$.
By replacing $r_k(\{\bQ_\ell\}_{\ell=1}^G)$ in constraint \eqref{eq.GRBrelaxed1const2} with
$\bar{r}_k(\{\bQ_\ell\}_{\ell=1}^G\mid\{\tilde{\bQ}_\ell\}_{\ell=1}^G)$
and by replacing \eqref{eq.GRBrelaxed1const1} with \eqref{eq.GRBrelaxed1const1a}, the problem in
\eqref{eq.GRBrelaxed1} can be approximated as
\begin{subequations}\label{eq.GRBrelaxed2}
\begin{align}
\max~ & \alpha \\
\text{subject to}~ & \left(\begin{array}{cc}R_g&\alpha\sqrt{\tau_g|\calK_g|}\\\alpha\sqrt{\tau_g|\calK_g|}&\sum_{k\in\calK_g}s_k\end{array}\right)\succeq\mathbf{0}, ~\forall g,\label{eq.GRBrelaxed2const1}\\
& \bar{r}_k(\{\bQ_\ell\}_{\ell=1}^G\!\mid\!\{\tilde{\bQ}_\ell\}_{\ell=1}^G)+\delta^{-1}(1\!-\!s_k)\!\geq\! R_g, ~ \forall k \in\calK_g,~ \forall g,\label{eq.GRBrelaxed2const2}\\
& 0\leq s_k \leq 1, ~\forall k\in\calK,~\eqref{eq.GRB2const1},\text{
and } \eqref{eq.GRB2const2},\\
\text{variables:}~ & \{\bQ_g\}_{g=1}^G,\{R_g\}_{g=1}^{G},\{s_k\}_{k\in\calK}, \alpha.\notag
\end{align}
\end{subequations}
The resulting optimization problem in \eqref{eq.GRBrelaxed2} is
convex and can be solved using general purpose solvers such as CVX
\cite{cvx}. Notice that this approximation is conservative in the
sense that solving \eqref{eq.GRBrelaxed2} yields a feasible
approximate solution to \eqref{eq.GRBrelaxed1}. However, the
approximation performance can be improved by iteratively applying
the same approximation to \eqref{eq.GRBrelaxed1}, with
$\{\tilde{\bQ}_\ell\}_{\ell=1}^G$ taken as the solution of
$\{\bQ_\ell\}_{\ell=1}^G$ obtained in the previous iteration. The
process can be repeated until the value of $\alpha$ converges. The
algorithm is summarized in Algorithm \ref{alg.successive}.

\begin{algorithm}[h!]
\caption{Relaxed GRB via Successive Convex
Approximation (SCA)}\label{alg.successive}
\begin{enumerate}
\item Initialize $\{\tilde{\bQ}_\ell\}_{\ell=1}^G$ by any feasible point of \eqref{eq.GRB2const1}, \eqref{eq.GRB2const2}, and set $\tilde\alpha=0$.
\item Solve problem \eqref{eq.GRBrelaxed2}, and denote the solution of $\bQ_\ell$ as $\bQ_\ell^*$ and the objective value as $\alpha^*$.
\item If $|\tilde\alpha-\alpha^*|\leq\epsilon$, then stop; else, set $\tilde{\bQ}_\ell\leftarrow\bQ_\ell^*$ for all $\ell$, $\tilde\alpha\leftarrow \alpha^*$, and go to Step $2$.
\end{enumerate}
\end{algorithm}

The convergence of Algorithm \ref{alg.successive} is described in the following proposition.

\vspace{.1cm}
\begin{proposition}\label{prop.successive}\it
Let
$\{\{\bQ_g^*\}_{g=1}^G,\{R_g^*\}_{g=1}^G,\{s_k^*\}_{k\in\calK},\alpha^*\}$
denote the solution of \eqref{eq.GRBrelaxed2}. Then, any limit point
of the iterates
$\{\{\bQ^*\}_{g=1}^G,\{R_g^*\}_{g=1}^G,\{s_k^*\}_{k\in\calK},\alpha^*\}$
generated by Algorithm \ref{alg.successive} is a stationary point of
problem \eqref{eq.GRBrelaxed1}.
\end{proposition}
\vspace{.1cm}

The proof is given in Appendix \ref{app.prop.successive}.
After obtaining the values of $s_k$  from the relaxed problem (and the SCA approach in Algorithm \ref{alg.successive}), the sequential deflation technique proposed in the previous section can then be utilized to obtain a solution for the user subsets $\{\calA_g\}_{g=1}^G$. Similarly, by employing the sequential deflation technique, the optimization problem in \eqref{eq.GRBrelaxed2} is solved $O(K)$ times, which is a significant improvement over the subset search with bisection. The performance of the two algorithms will be compared in Section \ref{sec.simulations}.

\subsection{Fairness of the GRB Problem in the Non-I.I.D. Case based on Channel Normalization}\label{sec.subsec.MGfair}

Notice that, similar to the single-group scenario, the GRB problem may also suffer fairness issues since the average group-rate is again used as the maximization criterion in each group. In this section, we extend the heuristic channel normalization technique, previously proposed in Section \ref{sec.subsec.SGfair}, to the multi-group scenario.

Specifically, suppose that the channel vector $\bh_{b,k}[n]$ between BS $b$ and user $k$ in block $n$ has entries that are i.i.d. $\calC\calN(0,d_{b,k}^{-\alpha})$, where $d_{b,k}$ is the distance between BS $b$ and user $k$ and $\alpha$ is the path loss exponent. Let $\bar d_{b,g}=\frac{1}{|\calK_g|}\sum_{k\in\calK_g} d_{b,k}$ be the average distance between BS $b$ and users in group $g$. Then, the normalized channel vector between BS $b$ and user $k$ in group $g$ can be defined as
\begin{equation}\label{eq.normalized_channel}
\tilde \bh_{b,k}[n]=\bh_{b,k}[n]\sqrt{\frac{{\bar d_{b,g}}^{-\alpha}}{d_{b,k}^{-\alpha}}}.
\end{equation}
Here, instead of scaling the normalized channel vectors by the average distance associated with all users, as in \eqref{eq.normalized_channel}, the channel vectors are scaled by the average distance of users within its own group. Similar to Section \ref{sec.subsec.SGfair}, the normalized channel vectors are then utilized to compute the optimal user subsets, denoted by $\tilde \calA_g^*$, for all $g$, using the algorithm proposed in the previous subsection. With user subsets $\{\tilde\calA^*\}_{g=1}^G$, the optimal transmit covariance matrix can then be computed by solving \eqref{eq.GRBwithA1} using the original channel vectors $\bh_{b,k}$, $\forall b,k$.

By performing OUS based on the normalized channel vectors, users in the same group will have equal probability of being selected in each block. These users will likely be those whose instantaneous channels are temporarily more favorable, either in the sense that their channel gains are large relative to their respective averages or in the sense that their channel directions cause less interference to other groups. The performance of the proposed fair OUS scheme is demonstrated through simulations in Section \ref{sec.simulations}.

\section{Group-Rate Balancing based on Partial Channel Feedback}\label{sec.feedback}

In previous sections, the design of the transmit covariance matrices and the
user subsets was performed based on instantaneous knowledge of
all users' CSI. In practice, this may require a large amount of
feedback, especially when the number of users in the system is
large. In this section, we consider the case where only a subset of
users feeds back their CSI in each block. Since the CSI of some
users may be missing, the optimization can only be
performed based on the expected group-rate, i.e., the expectation of the average group-rate with respect to the statistics of the unknown channels.

Let $\calK_{g,{\rm fb}}$ be the set of users from group $\calK_g$ that have chosen to feedback their channels in the given block and let $\calK_{g,{\rm fb}}^c=\calK_g-\calK_{g,{\rm fb}}$ be the complement set, i.e., the set of users from group $\calK_g$ whose CSI is unknown. In this case, the system parameters can be found by solving the group-rate balancing with partial feedback (GRB-PF) problem as given below:

\noindent\underline{\bf GRB with Partial Feedback (GRB-PF) Problem:}
\begin{subequations}\label{eq.GRB-PF1}
\begin{align}
\mathop{\max} ~  &\mathop{\min_{  g   }} \frac{R_g}{\tau_g|\calK_g|}\E\!\left[\sum_{k\in\calK_g}\!{\bf 1}_{\left\{ r_k(\{\bQ_\ell\}_{\ell=1}^G)\geq R_g\right\}}\right] \\
\text{subject to}~ &
\sum_{g=1}^G{\text{tr}(\{\bQ_g\}_{b,b})} \leq P_{b}, ~\forall b,~~  \bQ_g \succeq\mathbf{0}, \forall g,\label{eq.GRB-PF1const1}\\
& \{\bQ_g\}_{b,b'}\!=\!{\bf 0}_{M\!\times\! M}, \text{ for } b\notin\calB_g \text{ or }b'\notin\calB_g,\label{eq.GRB-PF1const2}\\
\text{variables:}~ & \{\bQ_g\}_{g=1}^G,\{R_g\}_{g=1}^{G},\notag
\end{align}
\end{subequations}
Here, we assume that a user's decision to feedback its channel (or not) is independent of its channel realizations. Otherwise, the expectation must be evaluated by conditioning on the specific feedback strategy\footnote{Note that, in cases with partial feedback (i.e., when only part of the users feedback their channels), the specific feedback strategy may have a significant impact on the group-rate performance.
In particular, allowing channel feedback from users with bad channels may improve service towards these users in each block but may limit the achievable multiuser and temporal diversity gains over time; and allowing feedback from users with good channels may have the opposite effect.
The optimal feedback strategy should exploit the tradeoff between these two effects, especially in the non-i.i.d. case, but requires further studies that go beyond the scope of this work. Readers are referred to \cite{huang_hwang_chen_2010} and \cite{li_wang_wang_zhou_2013} for further discussions on this topic.}.
In particular, the expectation term inside the objective function can be written as
\begin{align}
\sum_{k\in\calK_{g,{\rm fb}}}\!\!{\bf 1}_{\left\{ \log_2 \left(1+
\frac{\tr(\bQ_g\bh_k\bh_k^H)}{\sum_{\ell \neq g}\tr(\bQ_\ell\bh_k\bh_k^H)+1} \right)\geq R_g\right\}}\!+\!\!
\sum_{k\in\calK_{g,{\rm fb}}^c}\!\!\!
\Pr\left(\! \log_2 \!\left(\!1\!+\!
\frac{\tr(\bQ_g\bh_k\bh_k^H)}{\sum_{\ell \neq g}\tr(\bQ_\ell\bh_k\bh_k^H)\!+\!1} \right)\!\geq\! R_g\!\right).
\end{align}
Notice that the probability in the second term cannot be evaluated in closed-form and, thus, is difficult to handle from an optimization perspective.
To address this issue, we propose to consider an approximation technique where the probability  is replaced with its sample average approximation (SAA). Similar techniques have also been widely adopted in the area of stochastic optimization, e.g.,  \cite{wang_ahmed_2008, pagnoncelli_ahmed_shapiro_2009}.

More specifically, let $f_{\bh_k}$ be the density function of
$\bh_k$ and let $\{\bh_k^{(j)}\}_{j=1}^{J}$ be $J$ vectors randomly
generated according to $f_{\bh_k}$. Then, the probability can be
approximated as
\begin{align}
\Pr\!\left(\! \log_2 \!\left(\!1\!+\! \frac{\tr(\bQ_g\bh_k\bh_k^H)}{\sum_{\ell
\neq g}\!\tr(\bQ_\ell\bh_k\bh_k^H)\!+\!1}\! \right)\!\geq\! R_g\!\right)\!\approx
\frac{1}{J}\sum_{j=1}^J {\bf 1}_{\left\{ \log_2 \left(1+
\frac{\tr(\bQ_g\bh_k^{(j)}(\bh_k^{(j)})^H)}{\sum_{\ell \neq
g}\tr(\bQ_\ell\bh_k^{(j)}(\bh_k^{(j)})^H)+1} \right)\geq R_g\right\}}.
\end{align}
With knowledge of the distributions of the channel vectors $\bh_k$, for each
$k\in\calK^c_{g,\rm fb}$,
the objective function in
\eqref{eq.GRB-PF1} can be approximated as
\begin{align}
\min_{g}~\frac{R_g}{\tau_g|\calK_g|}&\left[
\sum_{k\in\calK_{g,{\rm fb}}}\!\!{\bf
1}_{\left\{r_k(\{\bQ_\ell\}_{\ell=1}^G)\geq R_g\right\}}+
\sum_{k\in\calK_{g,{\rm fb}}^c}\frac{1}{J}\sum_{j=1}^J {\bf
1}_{\left\{r_{k,j}(\{\bQ_\ell\}_{\ell=1}^G)\geq R_g\right\}}
\right],\label{eq.GRB-PFobj_approx}
\end{align}
where
\begin{equation}
r_{k,j}(\{\bQ_\ell\}_{\ell=1}^G)\triangleq\log_2\left(1+\frac{\tr(\bQ_g\bh_k^{(j)}(\bh_k^{(j)})^H)}{\sum_{\ell\neq{g}}\tr(\bQ_\ell\bh_k^{(j)}(\bh_k^{(j)})^H)+1}\right),
\end{equation}
for $k\in\calK^c_{g,\rm fb}$. {The channel vectors $\{\bh_k^{(j)}\}_{j=1}^J$ can be viewed as the channel vectors of $J$ virtual users associated with user $k$.  The entire set of virtual users in group $g$ is defined as $\calK_{g,{\rm vir}}\triangleq\calK_{g,{\rm fb}}^c\times\{1,\ldots, J\}$.
Moreover, let $\calA_{g,{\rm fb}}\subseteq\calK_{g,{\rm fb}}$ and $\calA_{g,{\rm vir}}\subseteq\calK_{g,{\rm vir}}$
be the subsets of selected feedback and virtual users, respectively, in group $g$ that are intended to successfully decode, and let the index pair $(k,j)$ denote the $j$-th virtual user associated with (non-feedback) user $k$. The GRB-PF problem in \eqref{eq.GRB-PF1} can then be approximated as follows:
\begin{subequations}\label{eq.GRB-PF-SAA0}
\begin{align}
\max~  &\mathop{\min_{  g \in \{ 1,...,G\}  }}  \frac{R_g}{\tau_g|\calK_g|}\left(|\calA_{g,{\rm fb}}|+\frac{1}{J}|\calA_{g,{\rm vir}}|\right)\\
\text{subject to}~ & r_k(\{\bQ_\ell\}_{\ell=1}^G)\geq R_g, ~\forall k\in\calA_{g,{\rm fb}}, \forall g,\\
& r_{k,j}(\{\bQ_\ell\}_{\ell=1}^G)\geq R_g, ~\forall (k,j)\in\calA_{g,{\rm vir}}, \forall g,\\
&\eqref{eq.GRB-PF1const1} \text{ and }
\eqref{eq.GRB-PF1const2},\\
\text{variables:}~ & \{\bQ_g\}_{g=1}^G,\{R_g\}_{g=1}^{G},\{\calA_{g,{\rm fb}}\}_{g=1}^G,\{\calA_{g,{\rm vir}}\}_{g=1}^G.\notag
\end{align}
\end{subequations}
Notice that the above problem is difficult to solve due to the combinatorial nature of the search over the user subsets $\{\calA_{g,{\rm fb}}\}_{g=1}^G$ and $\{\calA_{g,{\rm vir}}\}_{g=1}^G$. However, when $\{\calA_{g,{\rm fb}}\}_{g=1}^G$ and $\{\calA_{g,{\rm vir}}\}_{g=1}^G$ are given, the problem becomes the same as \eqref{eq.GRBwithA1} with
$\tau_g'\triangleq \tau_g|\calK_g|/(|\calA_{g,{\rm fb}}|\!+\!\frac{1}{J}|\calA_{g,{\rm vir}}|)$
and, thus, can be solved using the bisection search algorithm described in Algorithm \ref{alg.bisection}.

To determine the user selection (i.e., the choice of $\{\calA_{g,{\rm fb}}\}_{g=1}^G$ and $\{\calA_{g,{\rm vir}}\}_{g=1}^G$), we introduce, for each feedback user $k\in\calK_{g,\rm fb}$, a user selection variable
$s_k$ and, for each virtual user $(k,j)\in\calK_{g,{\rm vir}}$, a user selection variable $t_{k,j}$.}
Then, similar to the previous
sections, the approximated GRB-PF problem can be reformulated as
\begin{subequations}\label{eq.GRB-PF-SAA1}
\begin{align}
\mathop{\max}~& \mathop{\min_{g}}\frac{R_g}{\tau_g''}\left(\sum_{k\in\calK_{g,\mathrm{fb}}}\!\!\!s_k\!+\!\! \frac{1}{J}\sum_{(k,j)\in\calK_{g,\mathrm{vir}}}\!t_{k,j}\right)\\
\text{subject to}~& r_k(\{\bQ_\ell\}_{\ell=1}^G)+\delta^{-1}(1-s_k)\geq R_g,~\forall k\in\calK_{g,\rm fb}, ~\forall g,\\
& r_{k,j}(\{\bQ_\ell\}_{\ell=1}^G)+\delta^{-1}(1-t_{k,j})\geq R_g, ~\forall (k,j)\in\calK_{g,\mathrm{vir}}, \forall g,\\
&s_k\in\{0,1\},~\forall k\in\calK_{g,{\rm
fb}},\forall g,~\eqref{eq.GRB-PF1const1},
\eqref{eq.GRB-PF1const2},\\
&t_{k,j}\in\{0,1\}, ~ \forall (k,j)\in\calK_{g,\mathrm{vir}}, ~\forall{g},\\
\text{variables:}~ & \{\bQ_g\}_{g=1}^G,\{R_g\}_{g=1}^{G},\{s_k, \forall k\in\calK_{g,\mathrm{fb}}\}_{g=1}^G,\{t_{k,j}, \forall (k,j)\in\calK_{g,{\rm vir}}\}_{g=1}^G.\notag
\end{align}
\end{subequations}
where {$\tau_g''\triangleq \tau_g|\calK_g|$.}

By relaxing the integer constraints and by applying properties of the Schur complement, we can obtain a similar relaxed problem in its epigraph form, i.e.,
\begin{subequations}\label{eq.GRB-PF-SAArelaxed1}
\begin{align}
\max~ & \alpha \\
\text{subject to}~~&\!\!\left(\!\!\!\begin{array}{cc}R_g & \alpha\sqrt{\tau_g''}\\
\alpha\sqrt{\tau_g''} &
\sum\limits_{k\in\calK_g}\!\!s_k\!+\!\!\frac{1}{J}\!\sum_{(k,j)\in\calK_{g,\mathrm{vir}}}\!
t_{k,j}
\end{array}\!\!\!\right)\!\!\succeq\! {\bf 0},~\forall g,\label{eq.GRB-PF-SAArelaxed1const1}\\
&r_k(\{\bQ_\ell\}_{\ell=1}^G) +\delta^{-1}(1\!-\!s_k) \!\geq\! R_g,~ \forall k \in\calK_{g,\rm fb},~ \forall g,\label{eq.GRB-PF-SAArelaxed1const2}\\
& r_{k,j}(\{\bQ_\ell\}_{\ell=1}^G)\!+\!\delta^{-1}(1\!-\!t_{k,j})\geq R_g,~\forall (k,j)\in\calK_{g,\mathrm{vir}},~\forall{g},\label{eq.GRB-PF-SAArelaxed1const3}\\
&0\leq s_k \leq 1,~\forall k\in\calK_{g,{\rm fb}},~\eqref{eq.GRB-PF1const1}, \eqref{eq.GRB-PF1const2},\\
&0\leq
t_{k,j}\leq 1, ~\forall (k,j)\in\calK_{g,\mathrm{vir}},~\forall g,\\
\text{variables:}~ & \{\bQ_g\}_{g=1}^G,\{R_g\}_{g=1}^{G},\{s_k, \forall k\in\calK_{g,\mathrm{fb}}\}_{g=1}^G,\{t_{k,j}, \forall (k,j)\in\calK_{g,\mathrm{vir}}\}_{g=1}^G, \alpha.\notag
\end{align}
\end{subequations}
The above problem is still non-convex due to the
constraints in \eqref{eq.GRB-PF-SAArelaxed1const2} and
\eqref{eq.GRB-PF-SAArelaxed1const3}. Therefore, we adopt an SCA approach, similar to that in the previous section, where the left-hand-side of the inequalities in
\eqref{eq.GRB-PF-SAArelaxed1const2} and
\eqref{eq.GRB-PF-SAArelaxed1const3} are approximated by their
concave lower bounds. More specifically, the problem in
\eqref{eq.GRB-PF-SAArelaxed1} is approximated as
\begin{subequations}\label{eq.GRB-PF-SAArelaxed2}
\begin{align}
\max~~&\alpha \\
\text{subject to}~~&\!\!\left(\!\!\!\begin{array}{cc}R_g & \alpha\sqrt{\tau_g''}\\
\alpha\sqrt{\tau_g''} &
\sum\limits_{k\in\calK_g}\!\!s_k\!+\!\!\frac{1}{J}\!\sum_{(k,j)\in\calK_{g,\mathrm{vir}}}\!
t_{k,j}
\end{array}\!\!\!\right)\!\!\succeq\! \!{\bf 0},~\forall g,\label{eq.GRB-PF-SAArelaxed2const1}\\
&\bar{r}_k(\{\bQ_\ell\}_{\ell=1}^G\mid\{\tilde{\bQ}_\ell\}_{\ell=1}^G)+\delta^{-1}(1\!-\!s_k) \!\geq\! R_g,~ \forall k \in\calK_{g,\rm fb},~ \forall g,\label{eq.GRB-PF-SAArelaxed2const2}\\
&\bar{r}_{k,j}(\{\bQ_\ell\}_{\ell=1}^G\mid\{\tilde{\bQ}_\ell\}_{\ell=1}^G)+\!\delta^{-1}(1\!-\!t_{k,j})\geq R_g, ~\forall (k,j)\in\calK_{g,\mathrm{vir}},~\forall{g},\label{eq.GRB-PF-SAArelaxed2const3}\\
&0\leq s_k \leq 1,~\forall k\in\calK_{g,{\rm fb}},~\eqref{eq.GRB-PF1const1}, \eqref{eq.GRB-PF1const2},\\
&0\leq
t_{k,j}\leq 1, ~\forall (k,j)\in\calK_{g,\mathrm{vir}},~\forall{g},\\
\text{variables:}~ & \{\bQ_g\}_{g=1}^G,\{R_g\}_{g=1}^{G},\{s_k, \forall k\in\calK_{g,\mathrm{fb}}\}_{g=1}^G,\{t_{k,j}, \forall (k,j)\in\calK_{g,\mathrm{vir}}\}_{g=1}^G, \alpha,\notag
\end{align}
\end{subequations}
where $\{\tilde{\bQ}_\ell\}_{\ell=1}^G$ can be any point satisfying
constraints \eqref{eq.GRB-PF1const1} and \eqref{eq.GRB-PF1const2}, and
$\bar{r}_{k,j}(\{\bQ_\ell\}_{\ell=1}^G\mid\{\tilde{\bQ}_\ell\}_{\ell=1}^G)$
is defined as
\begin{align*}
\bar{r}_{k,j}(\{\bQ_\ell\}_{\ell=1}^G\mid\{\tilde{\bQ}_\ell\}_{\ell=1}^G)\triangleq& \log_2\!\left(\!1\!\!+\!\!\sum_{\ell=1}^G\!\tr(\bQ_\ell\bh_k^{\!(j)}(\bh_k^{\!(j)})^H)\!\right)-\!\log_2\!\bigg(\!1\!\!+\!\!\sum_{\ell{\ne}g}\tr(\tilde{\bQ}_\ell\bh_k^{\!(j)}{\bh_k^{\!(j)}}^{\!H})\!\bigg)\\
&-\!\frac{\sum_{\ell{\ne}g}\!\tr((\bQ_\ell\!-\!\tilde\bQ_\ell)\bh_k^{\!(j)}{\bh_k^{\!(j)}}^{\!H})}{\left[1\!+\!\!\sum_{\ell{\ne}g}\!\!\tr(\tilde{\bQ}_\ell\bh_k^{\!(j)}{\bh_k^{\!(j)}}^{\!H})\right]\ln 2}.
\end{align*}

The resulting optimization problem \eqref{eq.GRB-PF-SAArelaxed2} is
convex and can be solved using general purpose solvers such as CVX
\cite{cvx}. Similarly, refined approximate solutions can be obtained
by solving a series of convex optimization problems where, in each
iteration, $\{\tilde{\bQ}_\ell\}_{\ell=1}^G$ are taken as the
solutions of $\{\bQ_\ell\}_{\ell=1}^G$ obtained in the previous
iteration. The process can be repeated until the objective value
$\alpha$ converges. The algorithm is summarized in Algorithm
\ref{alg.successivePF}.

\begin{algorithm}[h!]
\caption{Relaxed GRB-PF via SAA and SCA}\label{alg.successivePF}
\begin{enumerate}
\item Randomly generate channel vectors $\{\bh_k^{(j)}\}_{j=1}^J, \forall k\in\calK_{g,{\rm fb}}^c,
    \forall g$.
\item Initialize $\{\tilde{\bQ}_\ell\}_{\ell=1}^G$ by any feasible point of \eqref{eq.GRB-PF1const1}, \eqref{eq.GRB-PF1const2}, and set $\tilde\alpha=0$.
\item Solve problem \eqref{eq.GRB-PF-SAArelaxed2}, and denote the solution of $\bQ_\ell$ as $\bQ_\ell^*$ , and the objective value as $\alpha^*$.
\item If $|\tilde\alpha-\alpha^*|\leq\epsilon$, then stop; else, set $\tilde{\bQ}_\ell\leftarrow\bQ_\ell^*$, $\tilde\alpha\leftarrow \alpha^*$, and go to Step 2.
\end{enumerate}
\end{algorithm}


\vspace{.1cm}
\begin{proposition}\label{prop.successivePF}\it
Let
$(\{\bQ_g^*\}_{g=1}^G,\{R_g^*\}_{g=1}^G,\alpha^*,\{s_k^*, \forall k\in\calK_{g,\mathrm{fb}}\}_{g=1}^G,\{t^*_{k,j}, {\forall (k,j)\in\calK_{g,\mathrm{vir}}}
\}_{g=1}^G)$
denote the solution of \eqref{eq.GRB-PF-SAArelaxed2}. Then, any
limit point of the iterates
generated by Algorithm \ref{alg.successivePF} is a stationary point
of problem \eqref{eq.GRB-PF-SAArelaxed1}.
\end{proposition}
\vspace{.1cm}

The proof of Proposition \ref{prop.successivePF} is similar to that
of Proposition \ref{prop.successive} and, thus, is omitted. After
obtaining the values of $s_k$ and $t_{k,j}$ from the relaxed
problem,  {a sequential deflation technique, similar to that in Algorithm \ref{alg.deflation}, is again applied to obtain a solution for the feedback and virtual user subsets, i.e., $\{\calA_{g,{\rm fb}}^*\}_{g=1}^G$ and $\{\calA_{g,{\rm vir}}^*\}_{g=1}^G$, respectively. The sequential deflation algorithm used for the GRB-PF problem is formally described in Algorithm \ref{alg.deflation_GRBPF}.
Notice that, in this problem, SAA introduces a large number of
virtual users which may increase the complexity of the sequential
deflation approach. To reduce complexity, we choose to eliminate more than one virtual user at once in each iteration. Specifically, if the
smallest user selection variable is associated with a virtual user
(i.e., if $\min_{(k,j)\in\calA_{g,{\rm vir}}} t_{k,j} \leq \min\{\min_{k\in\cup_g\calA_{g,{\rm fb}}} s_k,\min_{(k,j)\in\cup_g\calA_{g,{\rm vir}}} t_{k,j}\}$), then
$D_t> 1$ (e.g., $D_t=5$) virtual users with the $D_t$ smallest user selection variables
are eliminated from the  potential virtual user subset $\calA_{g,{\rm vir}}$. On the other hand, if $\min_{k\in\calA_{g,{\rm fb}}} s_k \leq \min\{\min_{k\in\cup_g\calA_{g,{\rm fb}}} s_k,\min_{(k,j)\in\cup_g\calA_{g,{\rm vir}}} t_{k,j}\}$, then $D_s=1$ feedback user is eliminated from the potential feedback user subset $\calA_{g,{\rm fb}}$. The removal of
virtual users can be viewed as the removal of channel realizations
that are not able to jointly support a high transmission rate.}

\begin{algorithm}[h!]
\caption{OUS by Sequential Deflation for the GRB-PF Problem}\label{alg.deflation_GRBPF}
    \begin{enumerate}
        \item[(i)] Initialize by setting $\calA_{g,{\rm fb}}\leftarrow\calK_{g,{\rm fb}}$, $\calA_{g,{\rm vir}}\leftarrow\calK_{g,{\rm vir}}$,
            $\calA_{g,{\rm fb}}^*\leftarrow\emptyset$, $\calA_{g,{\rm vir}}^*\leftarrow\emptyset$, for $g=1,\ldots,G$, and $\alpha^*\leftarrow0$.
        \item[(ii)] Solve \eqref{eq.GRB-PF-SAA0} for given user subsets $\{\calA_{g,{\rm fb}}\}_{g=1}^G$ and $\{\calA_{g,{\rm vir}}\}_{g=1}^G$, and let $\tilde \alpha$ be the resulting objective value.
        \item[(iii)] If $\tilde\alpha>\alpha^*$, then set $\alpha^*\leftarrow\tilde\alpha$,
            $\calA_{g,{\rm fb}}^*\leftarrow\calA_{g,{\rm fb}}$, and $\calA_{g,{\rm vir}}^*\leftarrow\calA_{g,{\rm vir}}$, $\forall g$.
        \item[(iv)] Solve the relaxed GRB-PF problem in \eqref{eq.GRB-PF-SAArelaxed2} (using Algorithm \ref{alg.successivePF}) for $\calK_{g,{\rm fb}}=\calA_{g,{\rm fb}}$ and $\calK_{g,{\rm vir}}=\calA_{g,{\rm vir}}$, $\forall g$, to yield the values of $s_k$, $\forall k\in\calA_{g,{\rm fb}}$, and $t_{k,j}$, $\forall (k,j)\in\calA_{g,{\rm vir}}$, $\forall g$.
        \item[(v)] If $\min_{k\in\calA_{g,{\rm fb}}} s_k \leq \min\left\{\min_{k\in\cup_g\calA_{g,{\rm fb}}} s_k,\right.$ $\left.\min_{(k,j)\in\cup_g\calA_{g,{\rm vir}}} t_{k,j}\right\}$, then set $\calA_{g,{\rm fb}}\leftarrow \calA_{g,{\rm fb}}-\calS_{\min,{\rm fb}}$, where $\calS_{\min,{\rm fb}}$ is the set of feedback users associated with the $D_s$ smallest values of $s_k$ among users in $\calA_{g,{\rm fb}}$; else, if $\min_{(k,j)\in\calA_{g,{\rm vir}}} t_{k,j} \leq \min\left\{\min_{k\in\cup_g\calA_{g,{\rm fb}}} s_k,\min_{(k,j)\in\cup_g\calA_{g,{\rm vir}}} t_{k,j}\right\}$, then          set $\calA_{g,{\rm vir}}\leftarrow \calA_{g,{\rm vir}}-\calS_{\min,{\rm vir}}$, where $\calS_{\min,{\rm vir}}$ is the set of virtual users associated with the $D_t$ smallest values of $t_{k,j}$ in $\calA_{g,{\rm vir}}$.
        \item[(vi)] Repeat steps (ii)-(v) until $\calA_{g,{\rm fb}}=\emptyset$ for some $g$. Then, take $\{\calA_{g,{\rm fb}}^*\}_{g=1}^G$ and $\{\calA_{g,{\rm vir}}^*\}_{g=1}^G$ as the desired set of feedback and virtual users.
    \end{enumerate}
\end{algorithm}

It is worthwhile to note that, in the proposed scheme, the virtual users are generated at random and, thus, the solution obtained under certain realizations of the virtual users' channels may be worse than the case without virtual users (i.e., solving the GRB problem assuming only the existence of the feedback users). This is especially the case when the number of virtual users, i.e., $J$, is small. Hence, in practice, one can take the better of the two schemes (i.e., the scheme considering only feedback users' channels and the proposed GRB-PF scheme) in each realization to guarantee that no loss occurs due to such randomness. This allows us to reduce the required number of virtual users in practice. Moreover, when the users' channels are i.i.d., the same set of virtual users can be used for all users, which further reduces the complexity.


\section{Simulations Results}\label{sec.simulations}

\begin{figure}[t]
     \centering
     \includegraphics[scale=.8]{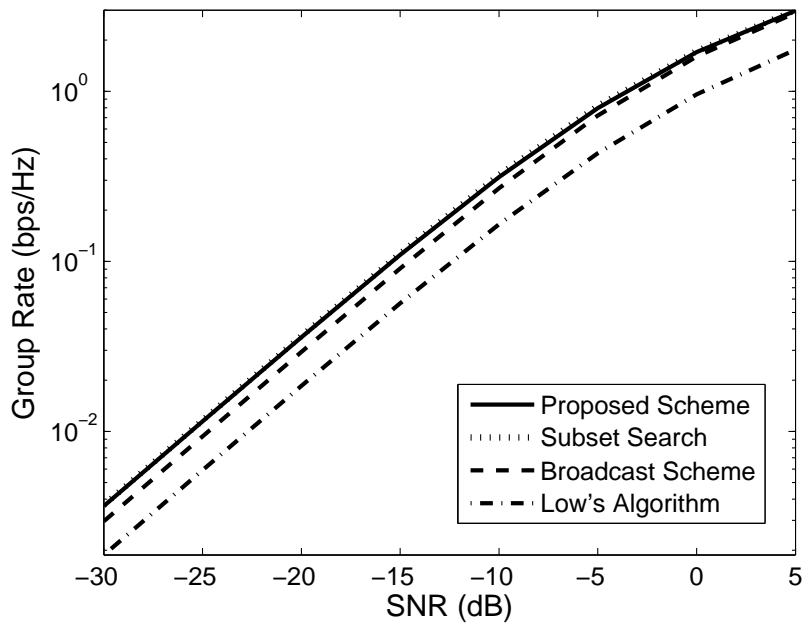}
     \vspace{-.2cm}
     \caption{For the single-group multicasting scenario with $B=3$ BSs, $M=2$ antennas per BS, and $K=30$ users, we show the average group-rates of single-group multicasting using the proposed sequential deflation, the subset search, the broadcast, and Low's algorithms.}
          \vspace{-.4cm}
     \label{fig.SG_GroupRate}
\end{figure}

\begin{figure}[t]
     \centering
     \includegraphics[scale=.8]{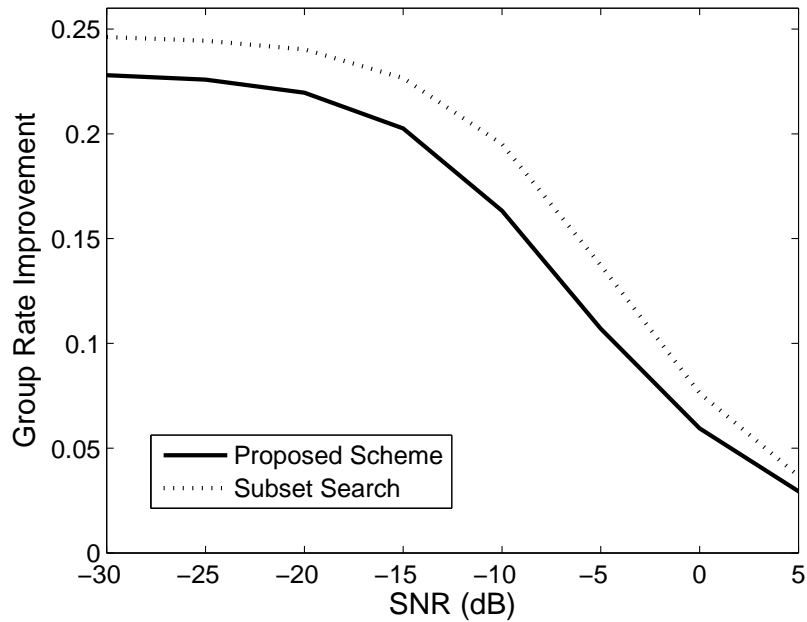}
     \vspace{-.2cm}
     \caption{For the single-group multicasting scenario with $B=3$ BSs, $M=2$  antennas per BS, and $K=30$ users, we show the average group-rate improvements obtained with the proposed sequential deflation algorithm and the subset search algorithm.}
          \vspace{-.4cm}
     \label{fig.SG_GRImprovement}
\end{figure}

\begin{figure}[t]
     \centering
     \includegraphics[scale=.8]{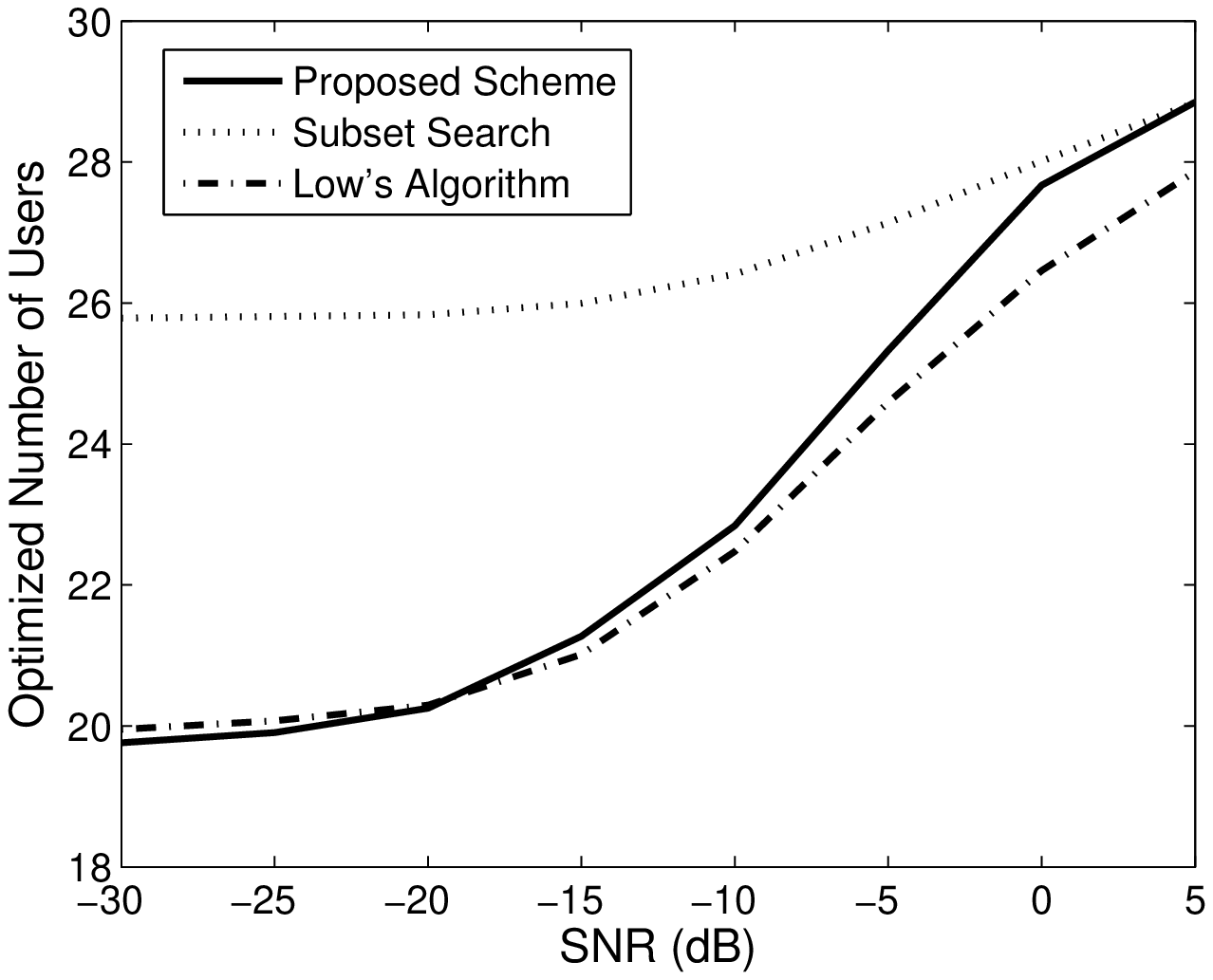}
     \vspace{-.2cm}
     \caption{For the single-group multicasting scenario with $B=3$ BSs, $M=2$ antennas per BS, and $K=30$ users, we show the average number of selected users obtained with the proposed sequential deflation, the subset search, and Low's algorithms.}
     \vspace{-.4cm}
          \label{fig.SG_OptUserNumber}
     \end{figure}

In this section, the effectiveness of the proposed schemes is demonstrated through computer simulations. In the experiments, we consider a multicell network with $B=3$ BSs, each equipped with $M=2$ transmit antennas, and set the power constraints of the BSs as $P_1=P_2=P_3=P$. The SNR is defined as $P/\sigma^2$, where $\sigma^2=1$ is the noise variance at each receiver (as chosen in Section \ref{sec.system_model}), and the entries of the channel vectors are assumed to be i.i.d. $\calC\calN(0,1)$ unless mentioned otherwise. The results are obtained by averaging over $600$ channel realizations.

\vspace{-.2cm}

\subsection{Single-Group Multicasting Scenario}

First, we consider the single-group multicasting scenario with $B=3$ BSs serving collaboratively $K=30$ users. We would like to emphasize that the scenario under consideration is different from having a single BS with $6$ antennas since each BS here is subject to their own individual power constraint $P$.
In Fig. \ref{fig.SG_GroupRate}, we compare the average group-rate of single-group multicasting using the proposed sequential deflation technique (described in Algorithm \ref{alg.deflation}) with that of single-group multicasting using the subset search algorithm (described in Algorithm \ref{alg.subset}), the broadcast scheme (where all users are served in each block), and Low's algorithm \cite{low_etal_2010} (which is based on a heuristic semi-orthogonal user selection algorithm). Here, Low's algorithm is performed with power allocation that takes into consider the individual power constraints at different BSs. We can see that, in the single group scenario, the proposed and the subset-search based OUS policies perform better than the broadcast scheme and, in fact, provide more advantages in the low SNR regime than in the high SNR regime. This is because, in the high SNR regime, a significant rate loss may be experienced when a user is eliminated and, thus, it is preferable to serve all users simultaneously (as done in the broadcast scheme). In cases with OUS, the subset search algorithm performs slightly better than the proposed scheme, but the difference is not significant and comes at the cost of much higher complexity. Low's algorithm performs the worst among all scehems since the precoder is not shaped in accordance with the individual power constraints, but chosen to maintain orthogonality among different signal directions \cite{low_etal_2010}. In Fig. \ref{fig.SG_GRImprovement}, we show the group-rate improvement of the proposed and the subset search algorithms. The group-rate improvement is defined as the difference in group-rate between the compared algorithm and the broadcast scheme, normalized by the group-rate of the latter scheme. We can see that, at low SNR, the group-rate improvement can be over $20\%$ for both the proposed and the subset search algorithms.

\begin{figure}[t]
     \centering
     \includegraphics[scale=.8]{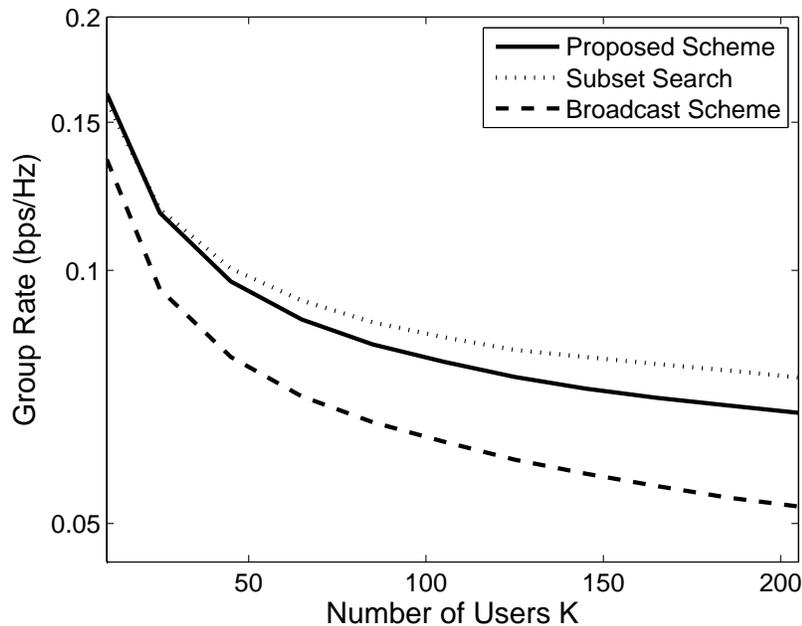}
     \vspace{-.2cm}
     \caption{For the single-group multicasting scenario with $B=3$ BSs, $M=2$ antennas per BS, and SNR=$-15$dB, we show the average group-rate of the proposed, the subset search, and the broadcast schemes as the number of users $K$ increases.}
          \vspace{-.4cm}
     \label{fig.Rev_SG_GroupRate_vs_K}
\end{figure}

In Fig. \ref{fig.SG_OptUserNumber}, we show the average number of users that are selected in each block in the proposed sequential deflation, the subset search, and Low's algorithms. We can see that the number of selected users increases with SNR in all schemes. The subset search algorithm eliminates fewer users because it terminates whenever no further improvement is obtained after removing a user whereas the proposed scheme first removes users sequentially until no user remains to yield $|\calK|$ candidate user subsets and then chooses the solution that yields the best group-rate. Even though the subset search algorithm is able to choose the most appropriate user to eliminate in each iteration (since it performs an exhaustive search among all remaining users in each iteration), it may have terminated prematurely because of the existence of many locally optimum solutions. It is worthwhile to note that the number of users selected in each block does not reflect the long-term fairness of the scheme. It only shows how each scheme exploits the tradeoff between multiuser diversity and multicast gains. When the users's channels are i.i.d., each user has equal opportunity of being selected in each block and thus, with the incorporation of the outer code, the average rate achieved by all users is asymptotically the same.

In Fig. \ref{fig.Rev_SG_GroupRate_vs_K}, we show the average group-rate of the proposed, the subset search, and the broadcast schemes with respect to the number of users, i.e., $K$. The receive SNR is fixed as $-15$dB. Due to the high computational complexity of the subset search scheme, its performance is averaged only over $300$ channel realizations for $K\leq 125$ and $30$ channel realizations for $K>125$. We can see that the OUS schemes (i.e., the proposed and the subset search algorithms) perform better than the broadcast scheme, especially as the number of users increases. This is because, in the broadcast scheme, the group-rate is limited by the worst user in the group and the channel conditions of the worst user will degrade continuously as the number of users increases. However, in the OUS schemes, the rate limitations caused by the worst users are alleviated by selecting users with sufficiently reliable channels in each block. For large $K$, the subset search algorithm performs better than the proposed scheme in the single-group scenario, but requires significantly higher computational complexity and rapidly becomes intractable as $K$ increases.


\begin{figure}[t]
     \centering
     \includegraphics[scale=.8]{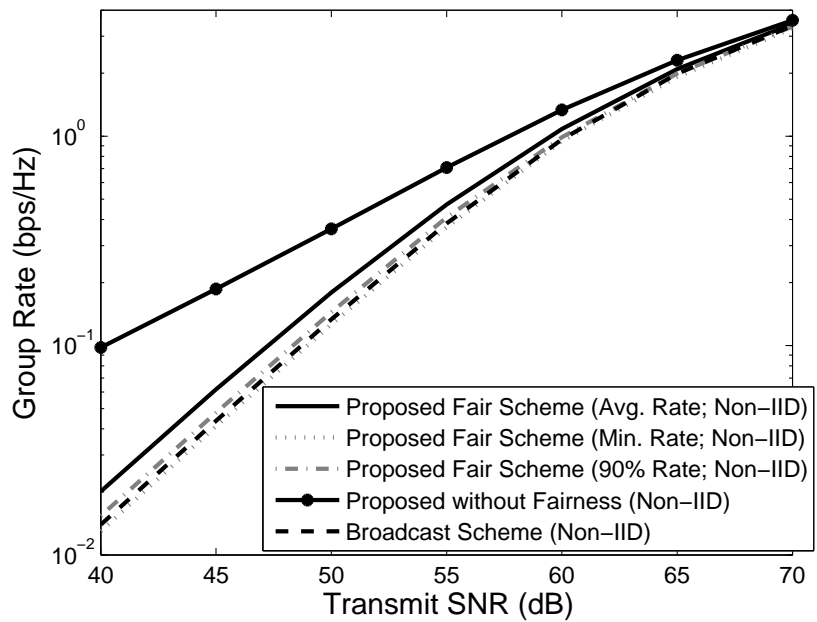}
          \vspace{-.2cm}
     \caption{For the single-group multicasting scenario with $B=3$ BSs, $M=2$ antennas per BS, and $K=30$ non-i.i.d. users, we show the average group-rates of single-group multicasting using the proposed OUS scheme with and without fairness considerations, and the broadcast scheme. The minimum and $90\%$ rates of the proposed fair OUS scheme is also shown for comparison.}
               \vspace{-.4cm}
     \label{fig.Rev_SG_NonIID_GroupRate_vs_TxSNR}
\end{figure}

Next, we consider the single-group multicasting scenario with $B=3$ BSs serving collaboratively $K=30$ users with non-identically distributed channel vectors. We assume that the users are uniformly distributed in a $[0,800]$m$\times [0,733]$m region and the coordinates of the $3$ BSs are $(150, 150)$, $(650, 150)$, and $(400,583)$, respectively. The BS locations are chosen such that they are distanced equally by $500$ meters.
The channel vector $\bh_{b,k}[n]$ is assumed to have entries that are i.i.d. $\calC\calN(0,d_{b,k}^{-\alpha})$ with path loss exponent $\alpha=2.5$. Here, $6$ sets of random user locations are considered, each averaged over $100$ channel realizations. In Fig. \ref{fig.Rev_SG_NonIID_GroupRate_vs_TxSNR}, we show the average group-rates versus the transmit SNR of the proposed single-group multicasting scheme with and without fairness considerations as well as that of the broadcast scheme. The minimum rate among all users (i.e., \eqref{eq.min_gropu_rate}) as well as the rate achieved by $90\%$ of users (called the $90\%$ rate) are also shown for comparison. The transmit SNR is given by the transmit power $P$ since the noise variance is set as $1$. The proposed scheme with fairness considerations refers to the scheme that utilizes normalized channel vectors to compute the user subsets whereas the scheme without fairness refers to the scheme that utilizes the original channel vectors to compute the user subsets. We can see that the proposed fair OUS scheme can still achieve average group-rate that is higher than the broadcast scheme even though a loss is experienced compared to the case without fairness. The minimum rate, however, can be lower than the rate achieved by the broadcast scheme if the message is not transmitted over a sufficiently large number of channel realizations. By averaging over $100$ channel realization in this figure, the minimum rate is slightly lower than the broadcast scheme due to the diversity of the rates among users. However, at least $90\%$ of users experience rates higher than that of the broadcast scheme, even though our scheme is derived using the average group-rate (instead of the minimum rate in  \eqref{eq.min_gropu_rate}) as the optimization criterion. This demonstrates the effectiveness of the proposed normalization. The minimum rate will become closer to the average rate as the transmission occurs over larger number of channel realizations.

\begin{figure}[t]
     \centering
     \includegraphics[scale=.8]{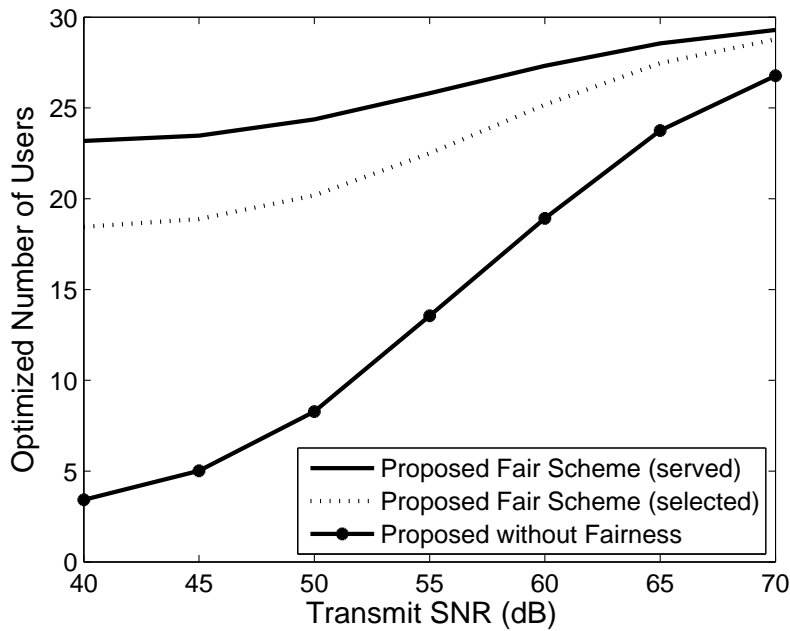}
               \vspace{-.2cm}
     \caption{For the single-group multicasting scenario with $B=3$ BSs, $M=2$ antennas per BS, and $K=30$ non-i.i.d. users, we show the average number of selected and served users obtained with the proposed OUS scheme with and without fairness considerations.}
               \vspace{-.4cm}
     \label{fig.Rev_SG_NonIID_SelectedUsers_vs_TxSNR}
\end{figure}

In Fig. \ref{fig.Rev_SG_NonIID_SelectedUsers_vs_TxSNR}, we show the average number of users that are selected in the schemes with and without fairness. We can see that, in the case without fairness considerations, less users are selected, and the selected users almost always correspond to users close to the BSs. As mentioned in Section \ref{sec.subsec.SGfair}, in the proposed fair OUS scheme, the users that are selected as target users are not the only users that may actually be served. In fact, by setting the rate to satisfy the worst user in the target user subset, users that are close to the BS (but outside of the target subset) may also have the opportunity to successfully decode. Hence, the number of users served is often greater than the number of users selected. However, we see from Fig.\,\ref{fig.Rev_SG_NonIID_SelectedUsers_vs_TxSNR} that there is not a significant difference between the two because of the directionality of the signal (i.e., the choice of the transmit covariance matrix). Interestingly, this is different from the single-antenna scenario where the two numbers may have a significant difference since, without spatial directionality, users close to the BS will have a high probability of being served when the target user subset includes a cell-edge user.

\vspace{-.2cm}

\subsection{Multi-Group Multicasting Scenario}

\begin{figure}[t]
     \centering
     \includegraphics[scale=.8]{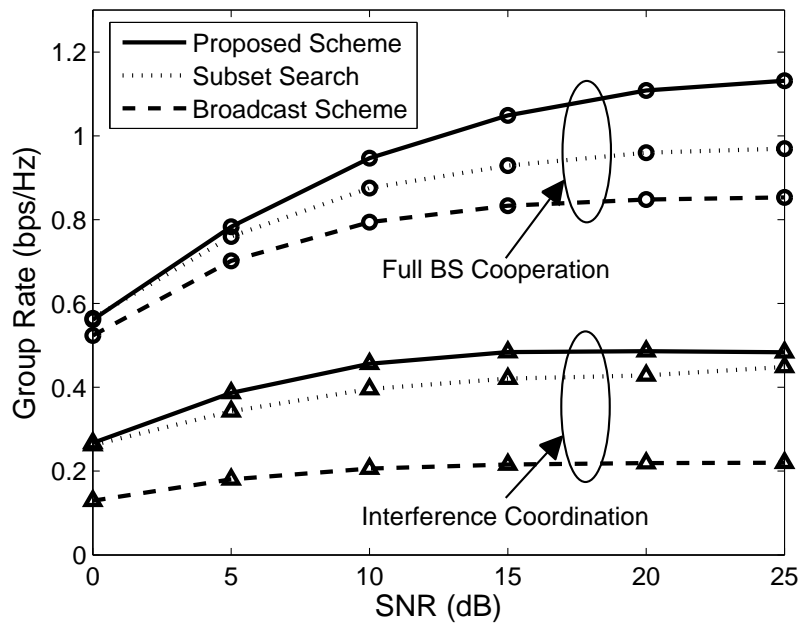}
          \vspace{-.2cm}
     \caption{For the multi-group multicasting scenario with $B=3$  BSs, $M=2$  antennas per BS, and $|\calK_1|=|\calK_2|=|\calK_3|=10$, we show the average group-rates of multi-group multicasting using the proposed, the subset search, the broadcast scheme. The results of both full BS cooperation and interference coordination scenarios are shown.}
          \vspace{-.4cm}
     \label{fig.MG_GroupRate}
\end{figure}


\begin{figure}[t]
     \centering
     \includegraphics[scale=.8]{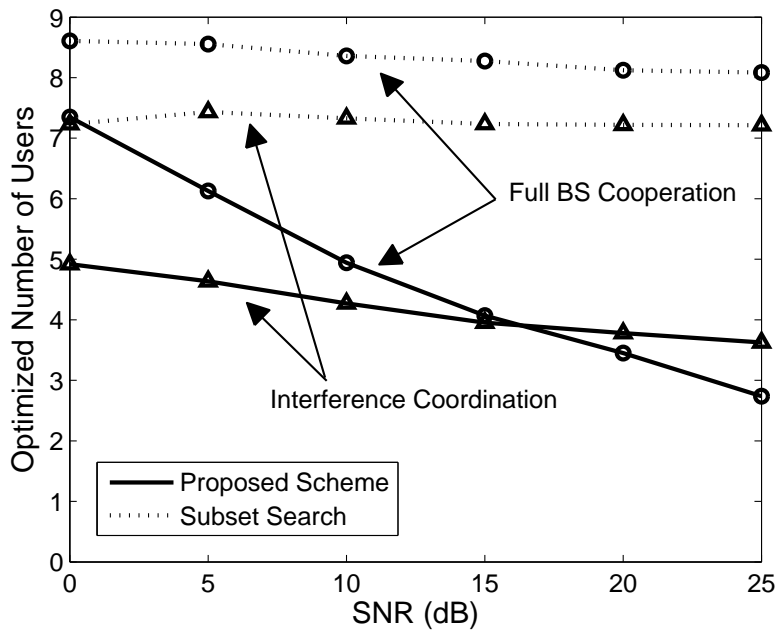}
          \vspace{-.2cm}
     \caption{For the multi-group multicasting scenario with $B=3$  BSs, $M=2$ antennas per BS, and $|\calK_1|=|\calK_2|=|\calK_3|=10$, we show the average number of selected users per group in the proposed and the subset search algorithms. The results of both full BS cooperation and interference coordination scenarios are shown.}
          \vspace{-.4cm}
     \label{fig.MG_OptUserNumber}
\end{figure}

In this section, we consider the multi-group multicasting scenario with $B=3$ BSs serving collaboratively $G=3$ multicast groups, each with $10$ users (i.e., $|\calK_1|=|\calK_2|=|\calK_3|=10$). In Fig. \ref{fig.MG_GroupRate}, we show the average group-rate of the relaxed GRB via SCA scheme proposed in Algorithm \ref{alg.successive} (c.f. Section \ref{sec.multiple_group}), and compare it with the subset search algorithm (averaged over only $300$ channel realizations) and the broadcast scheme, where all users are served simultaneously in each slot.
Both cases with full BS cooperation and interference coordination are considered. In the latter case, we assume that each BS serves only one group (namely, BS $b$ serves group $g$, for $b=g\in\{1,2,3\}$) and, thus, $\{\bQ_g\}_{b,b'}={\bf 0}_{M\times M}$, for all $b\neq b'$ and for all $b\neq g$. Different from the single-group scenario, we can see that user selection in the multi-group scenario is more advantageous in the high SNR regime and the gain is much more significant than that in the single-group case. This is because, in the multi-group scenario, the performance is interference limited at high SNR and, thus, user selection not only can help avoid rate limitations by the user with the worst channel but can also help reduce interference between signals intended for different groups. More interestingly, the proposed scheme also outperforms the subset search algorithm in the multigroup scenario since the latter scheme is more likely to converge towards a locally optimal solution in this scenario. These advantages can be observed in both full BS cooperation and interference coordination scenarios. 
Interestingly, the group-rate improvement is more significant for the case with only interference coordination. This is because, when BSs are not able to fully cooperate, the spatial degrees of freedom are not sufficient to effectively reduce interference solely through the design of the transmit covariance matrix. Therefore, the benefit of reducing interference through user selection is more pronounced in this case.

In Fig. \ref{fig.MG_OptUserNumber}, we show the average number of selected users per group when using the proposed algorithm under full BS cooperation and interference coordination. Interestingly, we can see that, different from the single-group scenario, the number of selected users is less at high SNR instead of at low SNR. This is due to the fact that, at high SNR, the performance is interference limited and, thus, the system would benefit more from eliminating users and reducing interference. This effect is more evident in the case of full BS cooperation where more spatial degrees of freedom are available for signal enhancement and interference avoidance.

\begin{figure}[t]
     \centering
     \includegraphics[scale=.8]{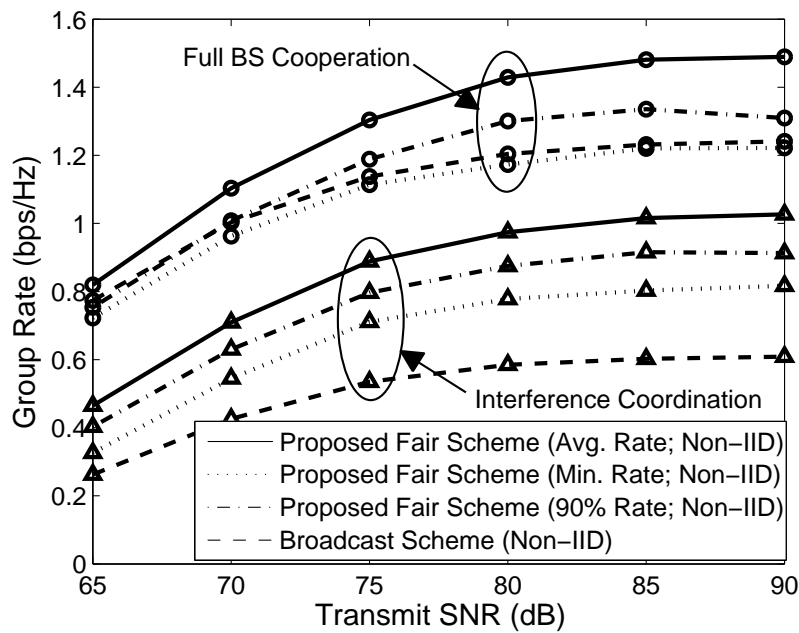}
          \vspace{-.2cm}
     \caption{For the multi-group multicasting scenario with $B=3$  BSs, $M=2$  antennas per BS, and $|\calK_1|=|\calK_2|=|\calK_3|=10$ non-i.i.d. users, we show the average group-rates of multi-group multicasting using the proposed fair OUS scheme and the broadcast scheme. The results of both full BS cooperation and interference coordination scenarios are shown. The minimum and $90\%$ rates of the proposed fair OUS scheme is also shown for comparison.}
          \vspace{-.4cm}
     \label{fig.Rev_MG_NonIID_GroupRate_vs_TxSNR}
\end{figure}

\begin{figure}[t]
     \centering
     \includegraphics[scale=.8]{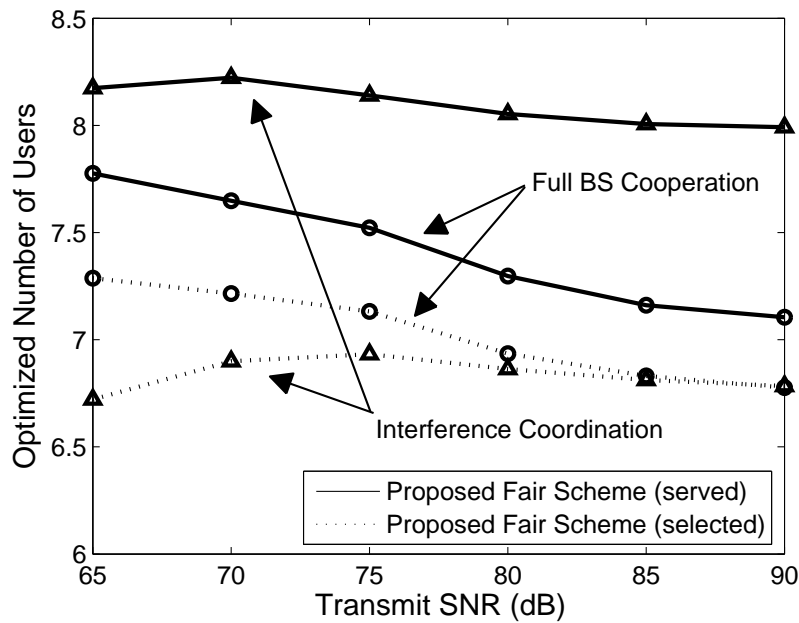}
          \vspace{-.2cm}
     \caption{For the multi-group multicasting scenario with $B=3$  BSs, $M=2$  antennas per BS, and $|\calK_1|=|\calK_2|=|\calK_3|=10$ non-i.i.d. users, we show the average number of selected and served users per group obtained with the proposed fair OUS scheme and the broadcast scheme. The results of both full BS cooperation and interference coordination scenarios are shown.}
          \vspace{-.4cm}
     \label{fig.Rev_MG_NonIID_SelectedUsers_vs_TxSNR}
\end{figure}

In Fig. \ref{fig.Rev_MG_NonIID_GroupRate_vs_TxSNR}, we consider the multi-group multicasting scenario with users whose channel vectors are non-identically distributed. Again, we have $B=3$ BSs serving collaboratively $G=3$ multicast groups, each with $10$ users (i.e., $|\calK_1|=|\calK_2|=|\calK_3|=10$). The users are deployed in the same way as that in Figs. 
\ref{fig.Rev_SG_NonIID_GroupRate_vs_TxSNR} and \ref{fig.Rev_SG_NonIID_SelectedUsers_vs_TxSNR}. Again, $6$ sets of random user locations are considered, each averaged over $100$ channel realizations.
Each BS serves a group consisting of the closest $10$ users. In the figure, we show the average group-rates of the proposed OUS scheme (c.f. Section \ref{sec.subsec.MGfair}) and the broadcast scheme. The minimum rate among all users and the $90\%$ rate are also shown for comparison. Both cases with full BS cooperation and interference coordination are considered. Similar to the i.i.d. case, we can see that the advantages of OUS increase with SNR and the gains are much more significant than the single-group scenario. Moreover, a significant advantage can still be observed in the case of interference coordination. This is because the normalized channel vectors used in the proposed OUS scheme preserves the direction of the channel vectors and, thus, is still able to successfully perform interference coordination among different BSs. Similar to the single-group scenario, the minimum rate may be smaller than the rate achieved in the broadcast scheme if the transmission does not occur over a large number of time slots, which is the case in the full BS cooperation scenario. However, the majority of users (in fact, over $90\%$ of users) achieve rates that are higher than that of the broadcast scheme. In Fig. \ref{fig.Rev_MG_NonIID_SelectedUsers_vs_TxSNR}, we show the average number of selected and served users obtained using the proposed fair OUS scheme under both full BS cooperation and interference coordination. We can see that the number of served users is again more than that of selected target users. However, this effect is less pronounced under full BS cooperation since the signal in this case contains directionality.

\begin{figure}[t]
     \centering
     \includegraphics[scale=.8]{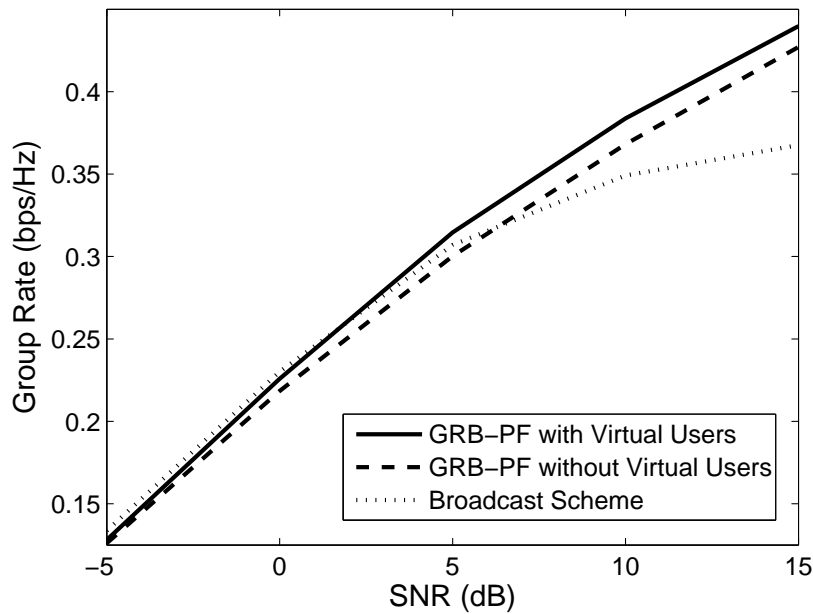}
          \vspace{-.2cm}
     \caption{Multi-group multicasting scenario with $B=3$ BSs, $M=2$ antennas per BS, and $|\calK_1|=|\calK_2|=|\calK_3|=20$, among which only $5$ per group feedback their channel vectors. We plot the average group-rate of the cases with and without virtual users.}
          \vspace{-.4cm}
     \label{fig.GRBPF_GroupRate}
\end{figure}

Finally, let us consider the GRB-PF problem as discussed in Section \ref{sec.feedback}. Here, we assume that there are $20$ users per group (i.e., $|\calK_1|=|\calK_2|=|\calK_3|=20$), but only $5$ per group feedback their channel vectors in each block. Notice that, in addition to the proposed scheme, it is also possible to compute the input covariance matrix, the rate, and the user selection assuming that only the users who feedback their CSI exist in the network. The latter is referred to as the case without virtual users. In the experiments, the proposed scheme is implemented with $J=100$ virtual users. In Fig. \ref{fig.GRBPF_GroupRate}, we show the average group-rate achieved when the system parameters are derived using the above two approaches. We can see that, by considering virtual users, the group-rate can be significantly improved, especially at high SNR where user selection is critical.
The performance of the broadcast scheme where the system parameters are designed by serving all feedback users in each block is also plotted for comparison. We can see that, at high SNR, where the performance is interference limited, user selection (even without consideration of virtual users) can provide significant group-rate improvement.

\section{Conclusion}\label{sec.conclusion}

In this work, the OUS scheme was examined for both single-group and multi-group multicasting scenarios in the physical layer of a multicell multi-antenna wireless system. User selection along with application layer erasure coding was proposed to overcome rate limitations caused by the worst user in the multicast group. The proposed user selection policies were derived based on the optimization and relaxation of a set of user selection variables. For the single group scenario, we formulated the problem as a group-rate maximization problem, and proposed an efficient user selection policy by performing a convex relaxation and by employing a sequential deflation algorithm. For the multi-group scenario, we formulated the problem as a group-rate balancing problem and proposed an efficient user selection policy by performing SCA along with the sequential deflation algorithm. When only part of the users feedback their instantaneous CSI, we further introduced the concept of virtual users to take into consideration the probability that non-feedback users are served in each block. The effectiveness of the proposed schemes was shown via computer simulations. Interestingly, we observed that user selection is more advantageous in the low SNR regime for the single-group scenario,  but is more beneficial in the high SNR regime for the multi-group scenario.


\appendices

\section{Proof of Lemma \ref{lemma.GRMequiv}}\label{app.lemma}

We first show that, for $\delta\leq
\left[\max_k\log_2(1+\sum_{b=1}^BP_b\|\bh_{b,k}\|^2)\right]^{-1}$,
 $(\bQ,R,\{s_k\}_{k\in\calK})$ is a
feasible point of \eqref{eq.GRM2} if and only if $(\bQ,R,\calA)$,
where $\calA=\{k\in\calK:s_k=1\}$, is a feasible point of
\eqref{eq.GRM1}. Specifically, let $(\bQ,R,\{s_k\}_{k\in\calK})$ be a
feasible point of \eqref{eq.GRM2} and let $\calA=\{k\in\calK:s_k=1\}$. We can see that, for any $k\in\calA$, the constraint in \eqref{eq.GRM1.const1} is equivalent to the constraint in \eqref{eq.GRM2.const1} when $s_k=1$. Therefore, if $(\bQ,R,\{s_k\}_{k\in\calK})$ is a
feasible point of \eqref{eq.GRM2}, then $(\bQ,R,\calA)$ must be a feasible point of \eqref{eq.GRM1.const1} as well. On the other hand, let $(\bQ,R,\calA)$ be a feasible point of \eqref{eq.GRM1.const1} and let $s_k=1$ if $k\in\calA$, and $s_k=0$, if $k\notin\calA$.
Similarly, the constraints in \eqref{eq.GRM2.const1} are the same as those in \eqref{eq.GRM1.const1}, for $k\in\calA$ (i.e., for $k$ such that $s_k=1$). However, for $k'\notin \calA$ (i.e., for $k'$ such that $s_{k'}=0$), the constraints in \eqref{eq.GRM2.const1} are redundant since
$\log_2[ 1+ \text{tr}(\bQ\mathbf{h}_{k'}\mathbf{h}_{k'}^{H})] +\delta^{-1}(1-s_{k'})\geq \log_2[ 1+ \text{tr}(\bQ\mathbf{h}_{k}\mathbf{h}_{k}^{H})]\geq R,$
for all $k\in\calA$. Therefore, $(\bQ,R,\{s_k\}_{k\in\calK})$ is a
feasible point of \eqref{eq.GRM2} if $(\bQ,R,\calA)$ is a feasible point of \eqref{eq.GRM1.const1}. Moreover, one can also see that $(\bQ,R,\calA)$ and $(\bQ,R,\{s_k\}_{k\in\calK})$ achieve the same objective values in their respective problems since
$\frac{1}{|\calK|}R\sum_{k\in\calK}s_k=\frac{1}{|\calK|}R\sum_{k\in\calK}\mathbf{1}_{\{k\in\calA\}}$.
The lemma follows.

\vspace{-.2cm}

\section{Proof of Proposition \ref{prop.successive}}\label{app.prop.successive}

We basically show that Algorithm \ref{alg.successive} is a special case of the successive upper-bound minimization (SUM) method in \cite{Razaviyayn_Hong_Luo_2013}.

First, notice that problem \eqref{eq.GRBrelaxed1} is the epigraph form of the following
problem:
\begin{subequations}\label{eq.GRBrelaxed3}
\begin{align}
\max~&\min_{g}\!\left\{\!\frac{\sum_{k\in\calK_g}\!\!s_k}{\tau_g|\calK_g|}\min_{k\in\calK_g}\!\left[r_k(\!\{\bQ_\ell\}_{\ell=1}^G)\!+\!\delta^{-1}(\!1\!-\!s_k\!)\right]\!\right\}~~\label{eq.GRBrelaxed3a}\\
\text{subject
to}~~&0{\le}s_k\le1,~\forall{k}\in\calK,~\eqref{eq.GRB2const1},~\text{and}~\eqref{eq.GRB2const2},\label{eq.GRBrelaxed3b}\\
\text{variables:}~& \{\bQ_g\}_{g=1}^G,\{s_k\}_{k\in\calK}.
\end{align}
\end{subequations}
Since $R_g$ and $\alpha$ are auxiliary variables, we can focus on
showing that any limit point of
$\{\{\bQ_g^*\}_{g=1}^G,\{s_k^*\}_{k=1}^K\}$ is a stationary point of
problem \eqref{eq.GRBrelaxed3}. Similar to problem
\eqref{eq.GRBrelaxed1}, problem \eqref{eq.GRBrelaxed2} is the
epigraph form of
\begin{subequations}
\begin{align}
\max~&\min_{g}\left\{\frac{\sum_{k\in\calK_g}s_k}{\tau_g|\calK_g|}\min_{k\in\calK_g}\!\left[\bar{r}_k(\{\bQ_\ell\}_{\ell=1}^G\!\mid\!\{\tilde{\bQ}_\ell\}_{\ell=1}^G)+\delta^{-1}(1-s_k)\right]\right\}\label{eq.GRBrelaxed4a}\\
\text{subject
to}~~&0{\le}s_k\le1,~\forall{k}\in\calK,~\eqref{eq.GRB2const1},~\text{and}~\eqref{eq.GRB2const2},\label{eq.GRBrelaxed4b}\\
\text{variables:}~& \{\bQ_g\}_{g=1}^G,\{s_k\}_{k\in\calK}.
\end{align}
\end{subequations}
The approximation
$r_k(\{\bQ_\ell\}_{\ell=1}^G)\approx\bar{r}_k(\{\bQ_\ell\}_{\ell=1}^G\mid\{\tilde{\bQ}_\ell\}_{\ell=1}^G)$
satisfies
$r_k(\{\tilde{\bQ}_\ell\}_{\ell=1}^G)=\bar{r}_k(\{\tilde{\bQ}_\ell\}_{\ell=1}^G\mid\{\tilde{\bQ}_\ell\}_{\ell=1}^G)$,
$r_k(\{\bQ_\ell\}_{\ell=1}^G)\ge\bar{r}_k(\{\bQ_\ell\}_{\ell=1}^G\mid\{\tilde{\bQ}_\ell\}_{\ell=1}^G)$, and
\[\left.\frac{\partial{r_k}(\{\bQ_\ell\}_{\ell=1}^G)}{\partial\bQ_g}\right|_{\bQ_\ell=\tilde{\bQ}_\ell,\forall\ell}=\left.\frac{\partial{r_k}(\{\bQ_\ell\}_{\ell=1}^G\mid\{\tilde{\bQ}_\ell\}_{\ell=1}^G)}{\partial\bQ_g}\right|_{\bQ_\ell=\tilde{\bQ}_\ell,\forall\ell},\]
for all $g$.
Therefore, Algorithm \ref{alg.successive} is essentially the SUM
method \cite{Razaviyayn_Hong_Luo_2013}. According to \cite[Theorem
1]{Razaviyayn_Hong_Luo_2013}, any limit point of
$\{\{\bQ_g^*\}_{g=1}^G,\{s_k^*\}_{k=1}^K\}$ generated by Algorithm
\ref{alg.successive} is a stationary point of
\eqref{eq.GRBrelaxed3}. Proposition \ref{prop.successive} is thus
proved.


{

}

\end{document}